\documentclass[a4paper,11pt]{article}
\usepackage{jcappub} 
\usepackage{lineno}

\arxivnumber{2303.14156} 
\title{\boldmath Non-Gaussianity in rapid-turn multi-field inflation}

 \author[a,b]{Oksana~Iarygina,}
 \author[b]{M.C.~David~Marsh,}
 \author[b]{Gustavo~Salinas}
 \affiliation[a]{Nordita, KTH Royal Institute of Technology and Stockholm University, \\Hannes Alfvéns väg 12, SE-106 91 Stockholm, Sweden}
 \affiliation[b]{The Oskar Klein Centre for Cosmoparticle Physics, Department of Physics,\\ Stockholm
University, AlbaNova, 10691 Stockholm, Sweden}

\emailAdd{oksana.iarygina@su.se, david.marsh@fysik.su.se, gustavo.salinas@fysik.su.se}

\abstract{We show that theories of inflation with multiple, rapidly turning fields can generate large amounts of non-Gaussianity.  We consider a general theory with two fields, an arbitrary field-space metric, and a potential that supports sustained, rapidly turning field trajectories. Our analysis accounts for non-zero field cross-correlation and does not fix the power spectra of curvature and isocurvature perturbations to be equal at horizon crossing. Using the $\delta N$ formalism, we derive a novel, analytical formula for bispectrum generated from multi-field mixing on super-horizon scales. 
Rapid-turn inflation can produce a bispectrum with several 
potentially large contributions that are not necessarily of the local shape.
 We exemplify the applicability of our 
 formula with a fully explicit model and show that the new contributions indeed can generate a large amplitude of local non-Gaussianity, $\fnl\sim {\cal O}(1)$. These results will be important when interpreting the outcomes of future observations.   }

\newcommand{\fnl}{\ensuremath{f_{\rm NL}^{\rm loc}}}

\makeatletter
\gdef\@fpheader{}
\makeatother

\begin{document}
\maketitle
\flushbottom

\section{Introduction}

Primordial non-Gaussianity is a powerful tool to discriminate between models of inflation by probing the dynamics and field content of the very early Universe. Single-field models of inflation most strongly couple momenta of similar wavelengths and result in bispectra that are highly suppressed in the `squeezed limit' where one long-wavelength-mode couple to two short-wavelength-modes. By contrast, multiple-field models of inflation and reheating can lead to couplings between different wavelengths and substantial signals in the squeezed limit of the bispectrum. Conventionally, this distinction can be quantified through the amplitude, \fnl, of the local shape function of the bispectrum. The parameter \fnl~is of particular interest as upcoming observations of the Cosmic Microwave Background (CMB) and the Large Scale Structure (LSS) aim to be sensitive to \fnl $\sim {\cal O}(1)$ \cite{Meerburg:2019qqi,Achucarro:2022qrl}. A detection of local non-Gaussianity with an amplitude of ${\cal O}(1)$ would rule out all attractor models of single-field inflation \cite{Maldacena:2002vr,Creminelli:2004yq}.

\emph{What would a hypothetical detection of \fnl $\sim O(1)$ tell us about multiple-field inflation?} Despite much progress on this question over the past two decades, the full answer remains elusive and a topic of active research. 
Several techniques have been used to calculate non-Gaussianities in multifield inflation, including the  $\delta N$-formalism \cite{ Sasaki:1995aw, Wands:2000dp,Lyth:2005fi} that captures non-Gaussianities generated from non-linear, classical evolution after horizon crossing and the in-in formalism that capture the non-linear effects at all scales \cite{Weinberg:2005vy}. More recently, interesting developments within the ``cosmological bootstrap" program, with a focus on massless exchanges during multi-field inflation (Ref.\cite{Wang:2022eop} and references therein) provide another model-independent way to compute bispectrum using a perturbative approach\footnote{See also novel ``cosmological flow" framework \cite{Werth:2023pfl}.}.

It is well established that large amplitudes of local non-Gaussianity can be generated through non-linear evolution close to the end of inflation or after it in spectator models, e.g.~the curvaton model \cite{Lyth:2001nq,Lyth:2002my,Bartolo:2003jx,Sasaki:2006kq}, hybrid and multi-brid inflation \cite{Lyth:2005qk,Alabidi:2006wa,Byrnes:2008zy,Sasaki:2008uc,Naruko:2008sq}, modulated and tachyonic
(p)reheating scenarios \cite{Dvali:2003ar,Enqvist:2004ey,Jokinen:2005by,Suyama:2007bg,Ichikawa:2008ne,Meyers:2013gua,Elliston:2014zea}. Generating substantial levels of non-Gaussianity \emph{during} a phase of multiple-field inflation is more challenging, as has been established in the simplest multi-field models  and using general arguments \cite{Seery:2005gb, Rigopoulos:2005ae,Bassett:2005xm,Rigopoulos:2005us,Kim:2006te,Battefeld:2006sz,Seery:2006js,Byrnes:2006vq,Yokoyama:2007dw,
Byrnes:2008wi,Byrnes:2009qy,Byrnes:2010em,Wands:2010af}, in studies of ultraviolet-motivated frameworks \cite{Renaux-Petel:2009jdf,Dias:2012nf,McAllister:2012am,Assassi:2013gxa,Ellis:2014gxa,Kawai:2014gqa,Kawai:2015ryj,Marzouk:2021tsz}  as well as using a statistical approach with random potentials \cite{Frazer:2011br,Dias:2017gva,Bjorkmo:2017nzd}. However, most studies to date have been limited to the slow-roll, slow-turn (SRST) approximation. In this approximation, elucidating analytical results \cite{Peterson:2010np, Peterson:2010mv, Peterson:2011yt} have been obtained for the amplitude \fnl~using the $\delta N$ and transfer function \cite{Amendola:2001ni,Wands:2002bn} formalisms: in terms of the tensor-to-scalar ratio $r$ and the isocurvature-to-curvature transfer function $T_{\cal R S}$, large amplitude of non-Gaussianities during SRST multiple field inflation can only be generated from the contribution
$$
\fnl \supset \frac{5 }{6} \sqrt{\frac{r}{8}} \left( \frac{T_{\cal RS}}{\sqrt{1 + T_{\cal RS}^2}} \right)^3 \partial_{\perp*} \ln T_{\cal RS} \, ,
$$
where $\partial_{\perp*}$ denotes a field-space derivative perpendicular to the direction of motion at horizon crossing (cf.~\cite{Bjorkmo:2017nzd} for similar expressions)\footnote{Throughout this paper we use natural units in which the reduced Planck mass is set to unity.}. Since $r$ is observationally constrained to be small and the cubed factor involving $T_{\cal RS}$ can never be large, it follows that generating $\fnl \sim {\cal O}(1)$ 
requires an exponential sensitivity of the transfer function to the precise trajectory at horizon crossing. Such a strong sensitivity is rare even in models with random potentials, hundreds of light fields, and appreciable amounts of isocurvature-to-curvature transfer \cite{Bjorkmo:2017nzd}\,\footnote{An example of sensitivity to initial conditions is a field trajectory that is rolling along a ridge in the potential, see, for instance, \cite{Lehners:2009ja}.}. 
This suggests that generating large \fnl~ in SRST multifield models requires some fine-tuning that can be difficult to realise.

However, over the past few years, it has increasingly been appreciated that multifield inflation need not rely on the slow-turn approximation. Indeed, in negatively curved field spaces, slow-turn trajectories can become unstable and transition into rapidly turning inflationary solutions \cite{Renaux-Petel:2015mga,Christodoulidis:2018qdw,Garcia-Saenz:2018ifx,Grocholski:2019mot}. Moreover, for certain field-space geometries, rapid-turn solutions can be realised without fine-tuning the inflationary potential \cite{Brown:2017osf,Mizuno:2017idt,Bjorkmo:2019aev,Bjorkmo:2019fls,Renaux-Petel:2021yxh} (which may be particularly interesting in the light of recent conjectures on the properties of effective potentials coming from ultraviolet completions \cite{Achucarro:2018vey} with recent developments \cite{Achucarro:2019pux,Achucarro:2019mea,Chakraborty:2019dfh,Aragam:2020uqi,Aragam:2021scu,Bhattacharya:2022fze,Anguelova:2022foz}). Non-Gaussianities arising at horizon crossing have been studied in detail in multi-field models with negatively curved field spaces  \cite{Garcia-Saenz:2018vqf,Fumagalli:2019noh,Bjorkmo:2019qno,Ferreira:2020qkf} and shown to peak near flattened triangle configurations with small contributions to \fnl \,\footnote{See also a recent review on non-Gaussianity in multi-field inflation with curved field space \cite{Garcia-Saenz:2019njm}. }. However, rapid-turn trajectories can lead to efficient conversion of isocurvature perturbations into adiabatic modes, thereby providing a source of non-Gaussianity during inflation captured by the $\delta N$ formalism.

In this paper we use the $\delta N$ formalism to derive a semi-analytical formula for non-Gaussianity in two-field models of inflation with sustained rapid turns. We identify new model-independent contributions to the non-Gaussianity parameter of the form
$$
\fnl \supset \frac{\eta_{\perp *}}{\sqrt{2\epsilon_*}}\, I_4+\tilde{M}_{\perp\perp *}\, I_5+ \tilde{M}_{\perp\parallel *}\, I_6 ,
$$
where $\eta_{\perp *}/\sqrt{2\epsilon_*}$ is the turn rate of field trajectory, $\tilde{M}_{ab *}$ is a parameter that encodes the background mass matrix and $I_i=I_i\left(T_{\cal RS},{\cal C}_{\cal RS}, P_{\cal S}/P_{\cal R}\right)$ are time-dependent coefficients that depend on the transfer function $T_{\cal RS}$, cross-correlation power spectrum ${\cal C}_{\cal RS}$ and ratio of isocurvature to curvature power spectra  $P_{\cal S}/P_{\cal R}$. We demonstrate that these new terms do not depend on initial conditions and give a significant contribution to the non-Gaussianity when the field trajectory is rapidly turning. It follows that large amplitudes, cf.~\fnl $\sim {\cal O}(1)$, can be generated  in rapid-turn models of inflation. The full bispectrum receives contributions from both sub-horizon and super-horizon evolution and need not peak in the local shape; still, large amplitude of local non-Gaussianities can be  observationally interesting even when the bispectrum receives significant contributions from other shapes.

The structure of the paper is as follows. In Section \ref{sec:BD} we introduce background equations of motion as well as other  background parameters relevant for further layout of the paper. In Section \ref{sec:Perturbations} we review the $\delta N$ formalism and apply it to compute two-point and three-point correlation functions in models of rapid-turn multi-field inflation. We discuss the general resulting formula for $f_{\rm NL}$, its non-locality and scale-dependence. In order to provide further analysis, we express the non-Gaussianity parameter via power spectrum at horizon crossing as well as at the end of inflation. Assuming scale-invariant power spectrum at horizon crossing, we compare our result with the one obtained before in the literature in SRST approximation. We further illustrate our findings with an example in Section \ref{sec:Example}.
We conclude in Section \ref{sec:Conclusion}.

\section{Background Dynamics}\label{sec:BD}

We investigate general two-field inflation models that are minimally coupled to gravity, described by the action of the form
\begin{equation}\label{action}
S= \int d^4 x \sqrt{-g}\left[ \frac{M_{\rm pl}^2}{2}R-\frac{1}{2}g^{\mu\nu}G_{ab}(\phi)\partial_{\mu}\phi^a \partial_{\nu}\phi^b-V(\phi^a)\right],
\end{equation}
where $g_{\mu\nu}$ is the spacetime metric, $R$ is the Ricci scalar constructed from spacetime quantities, $G_{ab}(\phi)$ is the field-space metric, and $V(\phi^a)$ is the multi-field potential. We use standard conventions, with Latin indices denoting scalar fields and Greek indices representing spacetime coordinates. 

Before discussing the different regimes of multi-field inflation, it is important to define certain quantities. An important parameter that controls the successful duration of inflation and encodes the deviation from the de Sitter expansion during inflation is the first slow-roll parameter, defined as
  \begin{equation}\label{epsilon}
 \epsilon = -\frac{H'}{H}=\frac{1}{2}G_{ab}\phi'^a  \phi'^b,
 \end{equation}
where $(')=d/dN=H^{-1} d/dt$ denotes a derivative with respect to the number of e-folds $N$ given by $dN=H dt$, with $H$ being the Hubble
parameter. The requirement $\epsilon\ll 1$ ensures a nearly exponential expansion during inflation. In order to have a prolonged quasi-de Sitter stage of inflation, $\epsilon$ needs to be small for a sufficient number of Hubble times. This happens when the second slow-roll parameter,  defined as
\begin{equation}
 \eta_H\equiv -\frac{\epsilon'}{2\epsilon},
 \end{equation}
is much smaller than one. Both conditions $\epsilon\ll 1, |\eta_H|\ll 1$ define the \textit{slow-roll approximation}.

The \textit{covariant acceleration} of the field
vector is given by
 \begin{equation}\label{eta}
 \eta^a = D_N \phi'^a,
 \end{equation}
where the covariant derivative on the field manifold with respect to the number of e-folds is defined as $D_{N}\stackrel{\text{def}}{=}\phi'^a \nabla_a$, with $\nabla_a$ being the covariant derivative in field space. The covariant derivative $D_N$ acts on an arbitrary field-space vector $A^a=A^a(\phi)$ as
\begin{equation}
D_{N} A^a \stackrel{\text{def}}{=} \partial_N \phi^b \nabla_b A^a = \partial_N A^a+\Gamma^a_{bc}\phi'^bA^c
\end{equation}
where $\Gamma^a_{bc}$ are the Christoffel symbols computed using $G_{ab}$.

To better understand the dynamics and perturbations in multi-field inflation it is useful to define unit vectors that are tangent and normal to the field trajectory. In the two-field case, these vectors, denoted by $e^a_{\parallel}$ and $e^a_{\perp}$ respectively, can be defined as follows \cite{GrootNibbelink:2000vx, GrootNibbelink:2001qt, Achucarro:2010da}
 \begin{gather}\label{kin_basis}
e^a_{\parallel}\equiv \frac{\phi'^a}{\sqrt{G_{bc}\phi'^b\phi'^c}}, \quad e^a_{\perp}\equiv s(N) \left( G_{bc} D_N e^b_{\parallel}  D_N e^c_{\parallel} \right)^{-1/2}D_N e^a_{\parallel},
 \end{gather}
where $s(N)=\pm 1$ and is introduced to avoid unphysical discontinuities\footnote{In the two-field case  the normal vector can be conveniently defined as $N_a\equiv \sqrt{{\rm det} G}\,  \epsilon_{ab} e^b_{\parallel}$, where $\epsilon_{ab}$ is the two-dimensional Levi-Civita symbol with $\epsilon_{11}=1$.}. We refer to this orthonormal basis as the \textit{kinematical basis}. The covariant acceleration $\eta^a$ can be expanded in this basis as
  \begin{equation}\label{eta_kin}
 \eta^a=\eta_{\parallel}e^a_{\parallel}+\eta_{\perp}e^a_{\perp}.
 \end{equation}
Note that the parallel component of the covariant acceleration is related to the slow-roll parameter $\eta_H$ as $\eta_\parallel = -\sqrt{2 \epsilon}\eta_H$. The definitions in equations \eqref{kin_basis} and \eqref{eta_kin} imply that rate of change of the tangent basis vector  is given by \cite{Peterson:2010np}
\begin{equation}
  e_{\perp a} D_N e^a_{\parallel}=\frac{\eta_{\perp}}{\sqrt{2\epsilon}}.
\end{equation}
Therefore, we refer to $\eta_{\perp}/\sqrt{2\epsilon}$ as the \textit{turn rate} parameter\footnote{It is worth noting that our definition of the turn rate parameter is related to the other commonly used definition of the turn rate $\Omega\equiv -e_{{\perp}a}D_t e^a_{\parallel}$ as $\Omega=-H\eta_{\perp}/\sqrt{2\epsilon}$.}. It shows how quickly the field trajectory is changing direction
along the field manifold and parameterizes the deviation of inflationary trajectory from a geodesic. Along a geodesic, $D_N e^a_{\parallel}=0$ and the turn rate is zero by definition.  We also define the \textit{speed up rate} as the combination $\eta_\parallel/\sqrt{2\epsilon}$, since it measures the logarithmic rate of change of the field speed $\eta_\parallel/\sqrt{2\epsilon} = (\log \epsilon)^\prime$. Given the definitions in equation \eqref{kin_basis} the turn rate is always positive (when non-zero), but the speed-up rate can be either positive or negative depending on whether the field speed is increasing or decreasing.
In terms of speed-up rate, the slow-roll approximation can be written as
\begin{equation}
    \epsilon \ll 1, \quad \left| \frac{\eta_{\parallel}}{\sqrt{2\epsilon}}\right|\ll 1.
\end{equation}

The \textit{slow-roll slow-turn} (SRST) approximation, which has been assumed in the majority of works on multi-field inflation, is defined as \cite{Peterson:2010np}
\begin{equation}
\epsilon \ll 1,\quad \left| \frac{\eta_{\parallel}}{\sqrt{2\epsilon}}\right|\ll 1, \quad \frac{\eta_{\perp}}{\sqrt{2\epsilon}}\ll 1,
\end{equation}
and is valid when the deviation of the inflationary trajectory from a geodesic is small. On the contrary, when deviation is large, one may define the \textit{slow-roll rapid-turn} (SRRT) approximation 
\begin{equation}\label{SRRTapprox}
\epsilon \ll 1,\quad \left| \frac{\eta_{\parallel}}{\sqrt{2\epsilon}}\right|\ll 1, \quad \frac{\eta_{\perp}}{\sqrt{2\epsilon}}\gg 1.
\end{equation}
A sustained rapid-turn regime requires the additional condition 
\begin{equation}\label{sustainedSRRT}
\frac{\eta_{\perp}'}{\eta_{\perp}}\ll 1.
\end{equation}
In this paper we only consider sustained rapid-turn inflation models, for which both equations \eqref{SRRTapprox} and \eqref{sustainedSRRT} are satisfied.

The background equation of motion for scalar fields derived from the action in Eq. \eqref{action} can be written in terms of the covariant acceleration $\eta^a$ and the first slow-roll parameter $\epsilon$ as
 \begin{equation}\label{backgroundeom}
 \frac{\eta^a}{(3-\epsilon)}+\phi^{\prime a} + \nabla^{a} \ln V=0.
 \end{equation}
 Using this equation and the definition of $\epsilon$ in Eq. \eqref{epsilon}, we find the relation
\begin{equation}\label{nablaEps}
\mathbf{\nabla}_a\epsilon= - \tilde{M}_{ab} \phi^{\prime b},
\end{equation}
where we have defined the \textit{`modified mass matrix'} as
\begin{equation}\label{Mtilde}
\tilde{M}_{ab}\stackrel{\rm def}{=}\cfrac{1}{1+ \frac{\eta_\parallel \sqrt{2\epsilon}}{(3-\epsilon)^2}}\left( M_{ab}+\frac{\nabla_a\eta_b}{(3-\epsilon)}\right),
\end{equation}
 with $M_{ab} = \nabla_a\nabla_b\ln V$ being the background mass matrix. It is worth noting that because of additional contribution coming from the covariant acceleration, the modified mass matrix $\tilde{M}_{ab}$ is not symmetric.
 For the forthcoming computations, it is convenient to express here the relation between $\tilde{M}^{ab}$ and $\eta^a$. To obtain this relation, we act with $D_{N}$ on the background equations of motion \eqref{backgroundeom} to find
\begin{equation}
 \eta^a=-\sqrt{2\epsilon}\,e^b_{\parallel}\tilde{M}_b{}^a.
\end{equation}
Therefore, projecting $\eta^a$ onto the kinematical basis vectors, we get
\begin{gather}\label{MtildeViaEta}
    \tilde{M}_{\parallel\parallel}=-\frac{\eta_{\parallel}}{\sqrt{2\epsilon}},\quad
    \tilde{M}_{\parallel\perp}=-\frac{\eta_{\perp}}{\sqrt{2\epsilon}}.
\end{gather}
The components of $\mathbf{\nabla}_a\epsilon$ can also be related to the covariant acceleration and to the modified mass matrix. From \eqref{nablaEps} and \eqref{MtildeViaEta} we obtain
\begin{equation}\label{epsParPerp}
(\nabla\epsilon)_{\parallel}=\eta_{\parallel}, \quad
  (\nabla\epsilon)_{\perp}=-\sqrt{2\epsilon}\,\tilde{M}_{\perp\parallel}~.
\end{equation}
Note that, in general, $\tilde{M}_{\perp\parallel}\neq \tilde{M}_{\parallel\perp}$.

\section{Analysis of Perturbations}\label{sec:Perturbations}

In this section, we express the curvature and isocurvature perturbations at the end of inflation in terms of these same quantities at horizon crossing using the $\delta N$ formalism \cite{ Sasaki:1995aw, Wands:2000dp,Lyth:2005fi}. This is then used to determine an analytical expression for the non-Gaussianity parameter $f_{\rm NL}$. In Section \ref{subsec:deltaN_review}, we briefly review the $\delta N$ formalism and use it in Section \ref{subsec:twopt} to obtain the power spectra of both curvature and isocurvature perturbations as well as the cross-correlation between the two as a function of the two-point correlation functions of field perturbations at horizon crossing. In Section \ref{subsec:bispectrum}, we perform a computation of the non-Gaussianity parameter, $f_{\rm NL}$,\footnote{From now on we drop the superscript `local' and refer to the non-Gaussianity parameter as $f_{\rm NL}$. }  in the case of large turn-rate, $\eta_{\perp}/\sqrt{2\epsilon} \gg 1$, and with non-zero cross-correlation at horizon crossing, ${\cal C}_{\cal RS*}\neq 0$. 
The resulting analytical formula for $f_{\rm NL}$ is the main result of this paper. We express it in terms of horizon-crossing quantities in Section \ref{subsec:SectionHorizon}, and in terms of quantities evaluated at the end of inflation in Section \ref{subsec:SectionEOI}. In Section \ref{subsec:ComparisonSRST}, we compare our result with the expression in the slow-roll slow-turn approximation.

\subsection{Review of $\delta N$ Formalism}\label{subsec:deltaN_review}

The $\delta N$ formalism is a powerful tool that allows for the calculation of curvature perturbations from inflation on super-horizon scales while avoiding the full machinery of higher-order perturbation theory. This approach is both simple and physically intuitive, making it an attractive option for understanding the dynamics of the Universe during inflation.

The curvature perturbation may be defined as a scalar perturbation
to the spatial metric for a given foliation of spacetime. The spatial metric $g_{ij}$ can be written as \cite{Lyth:2004gb,Lyth:2005du,Salopek:1990jq}
\begin{equation}
g_{ij} = a^2(t) \gamma_{ij}\, \mathrm{e}^{2\psi(t,\mathbf{x})}= \tilde{a}^2(t,\mathbf{x}) \gamma_{ij}~,
\end{equation}
where $a(t)$ is the scale factor, $\gamma_{ij}$ is a matrix with unit determinant, $\psi(t,\mathbf{x})$ is a perturbation and  $\tilde{a}(t,\mathbf{x})$ is the local scale factor that describes the expansion of the Universe including perturbations at each point in spacetime. 
The curvature perturbation on uniform density hypersurfaces, $\zeta(t, \mathbf{x})$, is defined on the time-slicing where the spacial hypersurfaces have uniform density, i.e. $\psi_{\rm UD}(t,\mathbf{x})\equiv \zeta (t,\mathbf{x})$. The local scale factor on the uniform density slice is then $\tilde{a}(t,\mathbf{x}) = a(t) \mathrm{e}^{\zeta(t,\mathbf{x})}$. Flat hypersurfaces are defined on flat time-slices with $\psi_{\rm flat}(t,\mathbf{x})=0$.

The basis of the $\delta N$ formalism is the counting of the number of e-folds during inflation in different local patches of the Universe \cite{Wands:2000dp}. This is done by considering the number of e-folds between an initial flat hypersurface $\Sigma_*$ and a final uniform density hypersurface $\Sigma$. The difference in the number of e-folds between different local patches of the Universe then gives the curvature perturbation on the final hypersurface.

The flat slice $\Sigma_*$ is defined at time $t_*$, which refers to the time when all the relevant modes have exited the horizon, i.e. time of horizon crossing. The amount of expansion (given in number of e-folds) from a point in $\Sigma_*$ to another point in the uniform energy density slice $\Sigma$ is then given by
\begin{equation}
    N(t, \mathbf{x}) = \ln \left(\frac{\tilde{a}(t,\mathbf{x})}{a(t_*)}\right)~,
\end{equation}
with $t$ being the time coordinate at the final slice. The $\delta N$ formalism then relates the amount of expansion from a point in $\Sigma_*$ to a point in $\Sigma$ to the curvature perturbation $\zeta(t, \mathbf{x})$ at the final slice $\Sigma$ as \cite{Sasaki:1995aw}
\begin{equation}\label{deltaN}
    \zeta(t, \mathbf{x}) = \delta N = N(t, \mathbf{x}) - N_0(t)~,
\end{equation}
where $N_0(t) = \ln[a(t)/a(t_*)]$ is the unperturbed number of e-folds, i.e., the amount of expansion from $\Sigma_*$ to a flat hypersurface $\Sigma^\prime$ that is tangent to $\Sigma$ at the point $(t, \mathbf{x})$.

The number of e-folds $N(t, \mathbf{x})$ is in general a function of the values of the field and field velocities at the initial slice $\Sigma_*$. Here, we assume that given a point in the uniform density slice $\Sigma$, there corresponds a unique point $(t_*, \mathbf{x}_*)$ in $\Sigma_*$ along the world-line of the cosmological fluid. Therefore, in general $\delta N \equiv \delta N (\phi^a(t_*, \mathbf{x}_*), \dot{\phi}^a(t_*, \mathbf{x}_*); t)$. 
Under certain assumptions, the field velocities can be expressed as functions of the values of the fields only. This is the case, for example, if the field trajectory follows a slow-roll attractor \cite{Sasaki:1995aw}. Another case of interest in which this happens is for rapid-turn attractors \cite{Bjorkmo:2019fls}. In those instances, the number of e-folds becomes a function in field space, instead of in phase space, and one has $\delta N \equiv N (\phi^a(t_*, \mathbf{x}_*); t)$. One can then expand it as series on the initial field perturbations at time $t_*$ as
\begin{equation}\label{deltaN_expansion}
    \delta N = N_{a} \delta \phi_*^a + \frac{1}{2} N_{ab} \delta \phi_*^a \delta \phi_*^b + ...
\end{equation}
with $N_a = \nabla_a N |_*$, $N_{ab} = \nabla_a\nabla_{b} N |_*$, where the covariant derivatives are computed 
with respect to the field values at horizon exit, and $\delta \phi^a(t_*, \mathbf{x}_*) \equiv \delta \phi_*^a$ are covariant gauge-invariant field perturbations described in detail in Appendix \ref{app:Covariant}.
In the next section we use the $\delta N$ formalism to compute two-point functions.

\subsection{Two-point functions}\label{subsec:twopt}

We can now use the $\delta N$ formalism to find the curvature and isocurvature power spectra and their correlated cross spectrum, defined in Fourier space as
\begin{align}\label{power_spectra}
    \begin{split}
        \langle \mathcal{R}_{\vec k_1} \mathcal{R}_{\vec k_2} \rangle &= (2\pi)^3 \delta^{(3)}(\vec{k}_1 + \vec{k}_2) P_{\mathcal{R}}(k_1)\\
        \langle \mathcal{R}_{\vec k_1} \mathcal{S}_{\vec k_2} \rangle &= (2\pi)^3 \delta^{(3)}(\vec{k}_1 + \vec{k}_2) C_{\mathcal{R}\mathcal{S}}(k_1)\\
        \langle \mathcal{S}_{\vec k_1} \mathcal{S}_{\vec k_2} \rangle &= (2\pi)^3 \delta^{(3)}(\vec{k}_1 + \vec{k}_2) P_{\mathcal{S}}(k_1)~,
    \end{split}
\end{align}
with $\mathcal{R}$ and $\mathcal{S}$ denoting the comoving curvature perturbation and the isocurvature perturbation, respectively. These are related to the field perturbations in the kinematical basis as \cite{Wands:2002bn}
\begin{equation}\label{RS_def}
    \mathcal{R} = \frac{\delta \phi_\parallel}{\sqrt{2\epsilon}} ~, ~~~~~~~~~~ \mathcal{S} = \frac{\delta \phi_\perp}{\sqrt{2\epsilon}}~.
\end{equation}
At superhorizon scales, $k \ll aH$, the comoving curvature perturbation coincides with the curvature at uniform-density slices \cite{Malik:2008im}, i.e., $\mathcal{R} \xrightarrow{k \ll aH} \zeta$, so we can use Eqs. \eqref{deltaN} and \eqref{deltaN_expansion} obtaining
\begin{equation}\label{2pt_R}
    \langle \mathcal{R}_{\vec k_1} \mathcal{R}_{\vec k_2} \rangle = \langle \zeta_{\vec k_1} \zeta_{\vec k_2} \rangle = N_a N_b \langle \delta {\phi_*^a}_{\vec k_1} \delta {\phi_*^b}_{\vec k_2} \rangle
\end{equation}
to lowest order in the field perturbations. The power spectrum of field perturbations $P^{*ab}_\phi$ at horizon crossing is defined as
\begin{equation}
    \langle \delta {\phi_*^a}_{\vec k_1} \delta {\phi_*^b}_{\vec k_2} \rangle = (2 \pi)^3 \delta^{(3)} (\vec k_1 + \vec k_2) P^{*ab}_\phi (k_1)~,
\end{equation}
which can be plugged into Eq. \eqref{2pt_R} along with the definition of the curvature power spectrum in Eq. \eqref{power_spectra} to give
\begin{equation}\label{PR_Pphi}
    P_\mathcal{R}(k) = N_a N_b P^{*ab}_\phi (k)~.
\end{equation}

We can relate the power spectra defined in Eq. \eqref{power_spectra} to the power spectrum of field perturbations using the relations in Eq. \eqref{RS_def} to obtain
\begin{equation}\label{PRS_Pphi}
\begin{split}
    \langle \mathcal{R}_{\vec k_1} \mathcal{R}_{\vec k_2} \rangle &= \frac{1}{2\epsilon} \langle \delta {\phi_\parallel}_{\vec k_1} \delta {\phi_\parallel}_{\vec k_2} \rangle ~\Rightarrow~ P_\mathcal{R*} = \frac{P_\phi^{*\parallel\parallel}}{2\epsilon_*}\\
    \langle \mathcal{R}_{\vec k_1} \mathcal{S}_{\vec k_2} \rangle &= \frac{1}{2\epsilon} \langle \delta {\phi_\parallel}_{\vec k_1} \delta {\phi_\perp}_{\vec k_2} \rangle ~\Rightarrow~ C_\mathcal{RS*} = \frac{P_\phi^{*\parallel\perp}}{2\epsilon_*}\\
    \langle \mathcal{S}_{\vec k_1} \mathcal{S}_{\vec k_2} \rangle &= \frac{1}{2\epsilon} \langle \delta {\phi_\perp}_{\vec k_1} \delta {\phi_\perp}_{\vec k_2} \rangle ~\Rightarrow~ P_\mathcal{S*} = \frac{P_\phi^{*\perp\perp}}{2\epsilon_*}
\end{split}
\end{equation}
where $P_\mathcal{R*}$, $C_\mathcal{RS*}$ and $P_\mathcal{S*}$ are the power spectra and cross-correlation at the time of horizon crossing $t_*$ and the projections of the power spectrum of field perturbations along the kinematical basis vectors are given by $P_\phi^{*\parallel\parallel} = {e_\parallel}_a {e_\parallel}_b P_\phi^{*ab}$, $P_\phi^{*\parallel\perp} = {e_\parallel}_a {e_\perp}_b P_\phi^{*ab}$ and $P_\phi^{*\perp\perp} = {e_\perp}_a {e_\perp}_b P_\phi^{*ab}$. Ultimately, we are interested in the curvature and isocurvature power spectra and cross-correlations at the end of inflation. These can be expressed in terms of the power spectra at horizon crossing and the so-called transfer functions, defined as \cite{Amendola:2001ni,Wands:2002bn}
\begin{equation}\label{transfer_funcs}
    \left(\begin{array}{c}
        \mathcal{R} \\
        \mathcal{S}
    \end{array}\right) = \left(\begin{array}{cc}
        1 & T_\mathcal{RS} \\
        0 & T_\mathcal{SS} 
    \end{array}\right) \left(\begin{array}{c}
        \mathcal{R}_* \\
        \mathcal{S}_* 
    \end{array}\right)
\end{equation}
with $T_\mathcal{RS}$ describing the effect of isocurvature perturbations feeding curvature perturbations and $T_\mathcal{SS}$ accounting for the decay/growth of isocurvature perturbations during the superhorizon evolution. Note that Eq. \eqref{transfer_funcs} reflects the fact that, on superhorizon scales, curvature perturbations are conserved in the absence of isocurvature perturbations and that curvature perturbations cannot feed isocurvature perturbations.  The transfer function $T_\mathcal{RS}$ is proportional to the turn rate of the field trajectory. When the turn rate is large, the sourcing of curvature by isocurvature perturbations becomes significant. It can be computed using the following expressions \cite{Amendola:2001ni,Wands:2002bn}
\begin{equation}\label{TRSdef}
    T_{\cal{R S}}(N_*,N)=\int_{N_*}^N dN' \gamma(N')T_{\cal S S}(N_*,N'), \quad T_{\cal{S S}}(N_*,N)={\rm exp}\left[ \int_{N_*}^N dN' \delta (N')  \right],
\end{equation}
where functions $\gamma$ and $\delta$ for general turning trajectory are defined in a slow-roll limit as \cite{Kaiser:2012ak, Christodoulidis:2018qdw}
\begin{equation}\label{gammaDeltaDef}
    \gamma=2\frac{\eta_{\perp}}{\sqrt{2\epsilon}}, \quad \delta =-2\epsilon -\frac{{\cal M}_{\perp\perp}}{V}+\frac{{\cal M}_{\parallel\parallel}}{V}-\frac{4 }{3 }\left( \frac{\eta_{\perp}}{\sqrt{2\epsilon}}\right)^2.
\end{equation}
Here ${\cal M}_{\perp\perp}=e_{\perp a}e_{\perp}^b{\cal M}^a{}_b$ and ${\cal M}_{\parallel\parallel}=e_{\parallel a}e_{\parallel}^b{\cal M}^a{}_b$ with
\begin{equation}\label{MTRS}
    {\cal M}^a{}_b=G^{ac}\nabla_b\nabla_c V- H^2 R^a_{d f b}\phi'^d\phi'^f,
\end{equation}
where $R^a_{dfb}$ is the Riemann tensor of the field-space manifold. The duration of sourcing of curvature perturbation by isocurvature depends on the evolution of isocurvature perturbation, which may be traced by the evolution of $T_{\cal{S S}}$. At first glance, a large turn rate would give rise to an exponential decay of isocurvature perturbation and hence a very limited time for sourcing. However, it is not always the case.  To provide more intuition, let us introduce the mass of isocurvature perturbation, $\mu^2$, that can be written as \cite{GrootNibbelink:2000vx, GrootNibbelink:2001qt}
\begin{equation}\label{mu}
    \mu^2=e_{\perp}^ae_{\perp}^b\nabla_a\nabla_b V+\epsilon H^2 \mathbb{R}+3 H^2\left( \frac{\eta_{\perp}}{\sqrt{2\epsilon}}\right)^2 = {\cal M}_{\perp\perp}+3 H^2\left( \frac{\eta_{\perp}}{\sqrt{2\epsilon}}\right)^2,
\end{equation}
where $\mathbb{R}$ is the Ricci scalar of the field-space metric. In the case of canonical kinetic terms and field spaces with positive scalar curvature, for large turn rates, the isocurvature fluctuation becomes heavy and may be integrated out. This allows for an effective single-field description of inflation \cite{Achucarro:2010da}, which leads to the non-Gaussianity of equilateral shape. However, this single-field effective theory does not hold for general rapid-turn models with curved field spaces. For instance, when the field-space metric is a hyperbolic manifold, the Ricci scalar of the field-space is negative,  and isocurvature mass $\mu^2$ may become negative even for large turn rates. This is the essence of geometrical destabilization of inflation \cite{Renaux-Petel:2015mga}. The exponential growth of the unstable fields can drive the system to new attractor solutions. One class of such models is called transiently tachyonic scenarios \cite{Garcia-Saenz:2018ifx} that can be described
by an effective single-field theory with an imaginary speed of sound around Hubble
crossing.
In this case, the bispectrum peaks in the folded shape \cite{Garcia-Saenz:2018ifx, Garcia-Saenz:2018vqf,Fumagalli:2019noh,Bjorkmo:2019qno,Ferreira:2020qkf}. Moreover, when all the contributions in equation \eqref{mu} cancel each other, the isocurvature perturbation becomes exactly massless. One example of this is the shift-symmetric orbital inflation model \cite{Achucarro:2019pux}, where the isocurvature modes are massless and freeze on
superhorizon scales, constantly sourcing the curvature perturbation and producing the non-Gaussianity of the local shape. 

It is important to note that the entropic perturbations are not guaranteed to be negligible at horizon crossing and can influence the bispectrum on superhorizon scales. In this case, single-field effective descriptions may capture the dominant contribution to the bispectrum but still miss contributions generated on superhorizon scales. To determine if this is the case, one must follow the evolution of the bispectrum, e.g.~through the $\delta N$-formalism until the entropic modes become sufficiently suppressed. 

In this work,
we derive a general formula for the bispectrum sourced on superhorizon scales in rapid-turn models without restrictive assumptions about the magnitude of the entropy mass. In Section \ref{sec:Example}, we will find that this general formula can capture substantial sourcing of the bispectrum even when the isocurvature perturbation decays a few e-folds after horizon crossing.

With the definition \eqref{transfer_funcs}, the two-point functions of perturbations at the end of inflation are then given by
\begin{equation}
\begin{split}
    \langle {\mathcal{R}}_{\vec k_1} {\mathcal{R}}_{\vec k_2} \rangle &= \langle {\mathcal{R}_*}_{\vec k_1} {\mathcal{R}_*}_{\vec k_2} \rangle + 2 T_\mathcal{RS} \langle {\mathcal{R}_*}_{\vec k_1} {\mathcal{S}_*}_{\vec k_2} \rangle + T_\mathcal{RS}^2 \langle {\mathcal{S}_*}_{\vec k_1} {\mathcal{S}_*}_{\vec k_2} \rangle \\
    \langle {\mathcal{R}}_{\vec k_1} {\mathcal{S}}_{\vec k_2} \rangle &= T_\mathcal{SS} \langle {\mathcal{R}_*}_{\vec k_1} {\mathcal{S}_*}_{\vec k_2} \rangle + T_\mathcal{RS} T_\mathcal{SS} \langle {\mathcal{S}_*}_{\vec k_1} {\mathcal{S}_*}_{\vec k_2} \rangle \\
    \langle {\mathcal{S}}_{\vec k_1} {\mathcal{S}}_{\vec k_2} \rangle &= T_\mathcal{SS}^2 \langle {\mathcal{S}_*}_{\vec k_1} {\mathcal{S}_*}_{\vec k_2} \rangle
\end{split}
\end{equation}
which implies that the curvature power spectra can be expressed as 
\begin{equation}\label{PRS_transfer}
    \begin{split}
        P_\mathcal{R} &= P_\mathcal{R*} + 2 T_\mathcal{RS} C_\mathcal{RS*} + T_\mathcal{RS}^2 P_\mathcal{S*} \\
        C_\mathcal{RS} &= T_\mathcal{SS} C_\mathcal{RS*} + T_\mathcal{RS} T_\mathcal{SS} P_\mathcal{S*} \\
        P_\mathcal{S} &= T_\mathcal{SS}^2 P_\mathcal{S*}~.
    \end{split}
\end{equation}

The curvature power spectrum given in Eq. \eqref{PRS_transfer} can then be expressed in terms of the power spectra of field perturbations using the result in Eq. \eqref{PRS_Pphi}, which gives 
\begin{equation}\label{PR_comparison1}
    P_\mathcal{R} = \frac{1}{2\epsilon_*} ( P_\phi^{*\parallel\parallel} + 2 T_\mathcal{RS} P_\phi^{*\parallel\perp} + T_\mathcal{RS}^2 P_\phi^{*\perp\perp} )~.
\end{equation}
At the same time, Eq. \eqref{PR_Pphi} can be used to obtain
\begin{equation}\label{PR_comparison2}
    P_\mathcal{R} = N_\parallel^2 P_\phi^{*\parallel\parallel} + 2 N_\parallel N_\perp P_\phi^{*\parallel\perp} + N_\perp^2 P_\phi^{*\perp\perp}
\end{equation}
where the components of the gradient of $N$ are defined by the relation
\begin{equation}\label{Na}
    N_a = N_\parallel {e_\parallel}_a + N_\perp {e_\perp}_a~.
\end{equation}
Comparing Eqs. \eqref{PR_comparison1} and \eqref{PR_comparison2} term by term, we get
\begin{equation}
    N_\parallel = \frac{1}{\sqrt{2\epsilon_*}} ~, ~~~~~~~~~~  N_\perp = \frac{T_\mathcal{RS}}{\sqrt{2\epsilon_*}}~.
\end{equation}
It is convenient to define a unit vector ${e_N}_a$ along the direction of $N_a$,
\begin{equation}\label{eN}
    {e_N}_a \stackrel{\rm def}{=} \cos \Delta_N {e_\parallel}_a + \sin \Delta_N {e_\perp}_a~,
\end{equation}
where $\Delta_N$ is \textit{the correlation angle}, which is defined \cite{Peterson:2010mv} by the relation
\begin{equation}\label{correlationAngle}
    \tan \Delta_N = T_\mathcal{RS}~.
\end{equation}
These definitions allow one to express the gradient of the number of e-folds as
\begin{equation}\label{Na_eN}
    N_a = \sqrt{\frac{1+T_\mathcal{RS}^2}{2\epsilon_*}} {e_N}_a = \frac{{e_N}_a}{\sqrt{2\epsilon_*} \cos \Delta_N}~.
\end{equation}

Using Eq. \eqref{Na_eN}, the covariant second derivatives of $N(t, \mathbf{x})$ can be written as 
\begin{equation}\label{Nab}
    N_{ab} = \nabla_{a} N_{b} |_* = \frac{\nabla_a e_{Nb} |_*}{2\sqrt{2\epsilon_*} \cos \Delta_N} + \frac{N_a}{2} \left(-\frac{1}{2\epsilon_*}\nabla_b \epsilon +\sin \Delta_N\cos \Delta_N \nabla_b T_{RS} \right)\biggr|_* ~.
\end{equation}
It is worth noting that in a computation of non-Gaussianity using the $\delta N$ formalism in the SRST approximation the first term in \eqref{Nab} projects out and does not contribute to the non-Gaussianity parameter. The reason is that it arises from the combination $N_a N_b N^{ab}$ and appears in the form $e_{Na} e_{Nb} \,\nabla^a e^b_N$ that gives zero as a product of two orthogonal vectors. However, when the cross-correlation spectrum is non-zero, it provides an additional contribution to the non-Gaussianity. We will show this in detail in the next section.

From the definition \eqref{eN} one can compute the gradient of the normal unit vector
\begin{equation}\label{nablaeN}
    \nabla_a e_{Nb} = \cos \Delta_N (\cos^2 \Delta_N \nabla_a T_{RS} + \theta_a) (e_{\perp b} - T_{RS} e_{\parallel b}),
\end{equation}
where in the last step we defined the vector $\theta^a$ as $\nabla_a {e_\parallel}_b \stackrel{\rm def}{=} \theta_a {e_\perp}_b$, which in turn implies $\nabla_a {e_\perp}_b = -\theta_a {e_\parallel}_b$. One may find that $\theta_a$ is given by
\begin{equation}\label{theta}
    \theta^a = \frac{\eta_{\perp}}{(3-\epsilon)^2}\tilde{M}^a{}_c e_\parallel^c -\left(\frac{1}{\sqrt{2\epsilon}}+\frac{\eta_{\parallel}}{(3-\epsilon)^2}\right)\tilde{M}^a{}_c e_\perp^c~.
\end{equation}
Let us  project now $\theta^a$ into tangent and normal directions. Using \eqref{nablaEps},\eqref{MtildeViaEta} and \eqref{epsParPerp} we find
\begin{gather}\label{thetaParPerp}
 \theta_{\parallel}=\frac{\eta_{\perp}}{2\epsilon}, \quad
  \theta_{\perp}=\frac{\eta_{\perp}}{(3-\epsilon)^2}\tilde{M}_{\perp\parallel}-\left(\frac{1}{\sqrt{2\epsilon}}+\frac{\eta_{\parallel}}{(3-\epsilon)^2}\right)\tilde{M}_{\perp\perp}.
\end{gather}
Having established the relations above, we can now proceed with the computation of the three-point functions.

\subsection{Bispectrum of curvature perturbations}\label{subsec:bispectrum}

In this section, we turn our attention to the calculation of the three-point functions of curvature perturbations. The bispectrum of the curvature perturbation $\zeta$ is defined via the three-point function
\begin{equation}\label{Bdef}
    \left\langle\zeta_{\vec{k}_{1}} \zeta_{\vec{k}_{2}} \zeta_{\vec{k}_{3}}\right\rangle = (2\pi)^3 \delta^{(3)} (\vec k_1 + \vec k_2 + \vec k_3) B_\zeta (k_1, k_2, k_3) ~.
\end{equation}

Particularly interesting is the \emph{`local'} bispectrum, which can be realised when the curvature perturbation 
can be written as a local function of position, i.e. locally in real space, in the form
\begin{equation}
    \zeta(x)=\zeta_g(x)-\frac{3}{5}f_{\rm NL}^{\rm loc}\left(\zeta_g(x)^2-\langle\zeta_g^2(x)\rangle \right),
    \label{eq:fnllocsinglefield}
\end{equation}
where \textit{the non-Gaussianity parameter} $f_{\rm NL}^{\rm loc}$ is a constant parameter that encodes the deviation from Gaussianity and $\zeta_g(x)$ being Gaussian. In this case the bispectrum is called \textit{`local'} as it appears to be of the \textit{`local form'}
\begin{equation}\label{Blocal}
    B_\zeta^{\rm loc}(k_1, k_2, k_3)= \frac{6}{5} f_{\rm NL}^{\rm loc}\left( P_\mathcal{R}(k_1) P_\mathcal{R}(k_2) + (\vec{k} {\rm \text { cyclic perms}})\right),
\end{equation}
where ($\vec{k}$ cyclic perms) represent terms with the momenta $\vec{k}_i$ permuted cyclically.
 In the case of scale-invariant  power spectra  the bispectrum further reduces to the \textit{`local shape'} 
 \begin{equation}
     B_\zeta^{\rm loc}(k_1, k_2, k_3) \propto f_{\rm NL}^{\rm loc} \left(\frac{1}{k_1^3 k_2^3} + (\vec{k} {\rm \text { cyclic perms}})\right).
 \end{equation}

 We note that \eqref{eq:fnllocsinglefield} is directly applicable to single-field models of inflation while in the rest of this paper, we use the general definitions \eqref{deltaN} and \eqref{deltaN_expansion}. Still, the local form of Eq.~\eqref{Blocal} is also most relevant for the super-horizon  contribution to the bispectrum in multi-field, slow-turn models of inflation. In rapid-turn models, as we will show,  
 the super-horizon contribution to the bispectrum takes a more general form that only reduces to the local shape under certain conditions.

 In general, the bispectrum can be defined as the sum over all shapes 
\begin{equation}
    B_\zeta(k_1, k_2, k_3)\propto \sum\limits_{\rm type}f^{\rm type}_{\rm NL} S_{\rm type}(k_1, k_2, k_3),
\end{equation}
with $S_{\rm type}(k_1, k_2, k_3)$ called \textit{the shape function}. The shape function sets the
 size of non-Gaussianity as a function of the triangle formed by the three-momenta $(\vec k_1, \vec k_2, \vec k_3)$ and encodes information about specific dynamics that produce non-Gaussianity \cite{Babich:2004gb}.

In this work we quantify the amount of non-Gaussianity by the parameter 
\begin{equation}\label{fNLdef}
    -\frac{6}{5} f_{\rm NL} \stackrel{\rm def}{=} \frac{B_\zeta(k_1, k_2, k_3)}{P_\mathcal{R}(k_1) P_\mathcal{R}(k_2) +(\vec{k} {\rm \text { cyclic perms}})}~
\end{equation}
and drop the superscript `loc' from now on to highlight that the parameter $f_{\rm NL}$ generally depends on details of the shape function and acquires a scale dependence. A scale dependence happens due to the non-linear evolution of initially Gaussian fluctuations after horizon crossing and/or when Gaussian fields constituting the system have different scale dependence \cite{Byrnes:2009pe}. When the non-Gaussianity is generated by local interactions on superhorizon scales, i.e. the bispectrum has a local form of Eq. \eqref{Blocal}, $f_{\rm NL}$ from Eq. \eqref{fNLdef} reduces to $f_{\rm NL}^{\rm loc}$ that is just a number.

To lowest order in the expansion on field perturbations, the three-point functions of curvature perturbations are given by
\begin{equation}\label{RRR3}
\left\langle\mathcal{R}_{\vec{k}_{1}} \mathcal{R}_{\vec{k}_{2}} \mathcal{R}_{\vec{k}_{3}}\right\rangle^{(3)} = \left\langle\zeta_{\vec{k}_{1}} \zeta_{\vec{k}_{2}} \zeta_{\vec{k}_{3}}\right\rangle^{(3)} = N_a N_b N_c \left\langle\delta {\phi_*^a}_{\vec{k}_{1}} \delta {\phi_*^b}_{\vec{k}_{2}} \delta {\phi_*^c}_{\vec{k}_{3}}\right\rangle~,
\end{equation}
where we used the $\delta N$ formalism expression given in Eq. \eqref{deltaN_expansion} to lowest order in the expansion. The three-point function $\left\langle\delta {\phi_*^a}_{\vec{k}_{1}} \delta {\phi_*^b}_{\vec{k}_{2}} \delta {\phi_*^c}_{\vec{k}_{3}}\right\rangle$ accounts for non-Gaussianities generated around horizon crossing, when some of the relevant modes were still evolving inside the horizon. In this work, we focus on the non-linear evolution of field perturbations on superhorizon scales and how it can lead to new types of non-Gaussianity in rapid-turn models. For that, we make use of the $\delta N$ Formalism, which only captures the non-linear classical evolution well after horizon crossing. The contribution from \eqref{RRR3}, on the other hand, depends on the quantities $\left\langle\delta {\phi_*^a}_{\vec{k}_{1}} \delta {\phi_*^b}_{\vec{k}_{2}} \delta {\phi_*^c}_{\vec{k}_{3}}\right\rangle$, which are determined by the subhorizon evolution of perturbations. Hence, the computation of the non-Gaussianity arising from the contribution described in Eq. \eqref{RRR3} is beyond the scope of this study. We will, however, get back to discussing the contribution from this term later in Sec. \ref{sec:Example}.

The next-to-leading order term is given by
\begin{equation}\label{RRR4}
\left\langle\mathcal{R}_{\vec{k}_{1}} \mathcal{R}_{\vec{k}_{2}} \mathcal{R}_{\vec{k}_{3}}\right\rangle^{(4)} = \frac{1}{2} N_a N_b N_{cd} \left\langle\delta {\phi_*^a}_{\vec{k}_{1}} \delta {\phi_*^b}_{\vec{k}_{2}} \left(\delta {\phi_*^c} \ast \delta {\phi_*^d}\right)_{\vec{k}_{3}}\right\rangle +(\vec{k} {\rm \text { cyclic perms}})~,
\end{equation}
where $\ast$ denotes a convolution, i.e.,
\begin{equation}
    \left(\delta {\phi_*^c} \ast \delta {\phi_*^d}\right)_{\vec{k}_{3}} = \int \frac{d^3k}{(2\pi)^3} \delta {\phi_*^c}_{\vec k} \delta {\phi_*^d}_{\vec k_3 - \vec k}~.
\end{equation}
One can use Wick's theorem to obtain
\begin{equation}
\begin{aligned}
    \left\langle\delta {\phi_*^a}_{\vec{k}_{1}} \delta {\phi_*^b}_{\vec{k}_{2}} \left(\delta {\phi_*^c} \ast \delta {\phi_*^d}\right)_{\vec{k}_{3}}\right\rangle &=\\ (2\pi)^3 \delta^{(3)} (\vec{k}_1 + \vec{k}_2 +& \vec{k}_3) N_a N_b N_{cd} \left[P_\phi^{*a c}\left(k_{1}\right) P_\phi^{*b d}\left(k_{2}\right)+(\vec{k} {\rm \text { cyclic perms}})\right]~,
    \end{aligned}
\end{equation}
which gives a bispectrum of the local form
\begin{equation}
    B_\zeta^{(4)}(k_1, k_2, k_3) = N_a N_b N_{cd} \left[P_\phi^{*a c}\left(k_{1}\right) P_\phi^{*b d}\left(k_{2}\right)+(\vec{k} {\rm \text { cyclic perms}})\right]~.
\end{equation}
Taking $B_\zeta(k_1, k_2, k_3) = B^{(3)}_\zeta(k_1, k_2, k_3) + B^{(4)}_\zeta(k_1, k_2, k_3)$, with $B^{(3)}_\zeta(k_1, k_2, k_3)$ defined by Eq. \eqref{Bdef}, but with the three-point function replaced by the lowest order contribution in Eq. \eqref{RRR3}. We then use Eq. \eqref{PR_Pphi} to get 
\begin{equation}\label{fNL34}
    -\frac{6}{5} f_{\rm NL} =  -\frac{6}{5} f^{(3)}_{\rm NL} - \frac{6}{5} f^{(4)}_{\rm NL}
\end{equation}
with
\begin{equation}\label{fNLGeneral}
    -\frac{6}{5} f^{(4)}_{\rm NL} = \frac{N_a N_b N_{cd} \left[P_\phi^{*ac}\left(k_{1}\right) P_\phi^{*b d}\left(k_{2}\right)+(\vec{k} {\rm \text { cyclic perms}})\right]}{N_e N_f N_g N_h \left[P_\phi^{*ef}(k_1) P_\phi^{*gh}(k_2) +(\vec{k} {\rm \text { cyclic perms}})\right]} \stackrel{\rm def}{=} \frac{N_a N_b N_{cd} K^{abcd}(k_1, k_2, k_3)}{N_e N_f N_g N_h K^{efgh}(k_1, k_2, k_3)}~,
\end{equation}
where we defined 
\begin{equation}\label{Kabcd}
    K^{abcd}(k_1, k_2, k_3) = P_\phi^{*ac}(k_1) P_\phi^{*bd}(k_2) + (\vec{k} {\rm \text { cyclic perms}}),
\end{equation}
and $f^{(3)}_{\rm NL}$ is given by Eq. \eqref{fNLdef}, but with $B^{(3)}_\zeta(k_1, k_2, k_3)$ instead of $B_\zeta(k_1, k_2, k_3)$.

Note that the quantity $K^{abcd}(k_1, k_2, k_3)$ defined in Eq. \eqref{Kabcd} only depends on the two-point functions of field perturbations at the time of horizon crossing $t_*$. We can consider these quantities as initial conditions to the subsequent evolution with all modes outside of the horizon. These quantities can be calculated numerically by tracking the evolution of modes inside the horizon and during horizon exit, and in some cases it is possible to determine $K^{abcd}(k_1, k_2, k_3)$ analytically. The clearest example of that is for the case of two light non-interacting fields. In that case, $P^{*ab}_\phi(k) = P_\phi^*(k) G^{ab}$ with the two-point function of massless scalar fields $P_\phi^*(k) = H^2_*/2k^3$ and one gets
\begin{equation}\label{Ksimple}
    K^{abcd}(k_1, k_2, k_3) = G^{ac} G^{bd} \left(P_\phi^*(k_1) P_\phi^*(k_2) + (\vec{k} {\rm \text { cyclic perms}})\right) ~\Rightarrow~ -\frac{6}{5} f_{\rm NL}^{(4)} = \frac{N_a N_b N^{ab}}{(N_c N^c)^2}~,
\end{equation}
which is an expression for the parameter $f_{\rm NL}$ that is widely used in the literature.

The expression in Eq. \eqref{fNLGeneral}, however, being more general, can account for cases in which the fields have arbitrary masses and non-trivial interactions during horizon crossing, which is not captured by the simple expression in Eq. \eqref{Ksimple}. Indeed, it follows the non-Gaussianity parameter in Eq. \eqref{fNLGeneral} is in general depends on the shape of the triangle $(\vec k_1, \vec k_2, \vec k_3)$ and acquires a scale dependence. We will discuss some of these aspects in the following sections. 

Mor intuitively, the quantity $K^{abcd}(k_1, k_2, k_3)$ can be written in the kinematical basis. Using the relations in Eq. \eqref{PRS_Pphi}, one may find 
\begin{equation}
\begin{aligned}
    K^{\parallel\parallel\parallel\parallel}(k_1, k_2, k_3) &= (2\epsilon_*)^2 \left(P_{\cal R *}(k_1)P_{\cal R *}(k_2) + (\vec{k} {\rm \text { cyclic perms}})\right),\\
    K^{\parallel\parallel\parallel\perp}(k_1, k_2, k_3) &= (2\epsilon_*)^2 \left(P_{\cal R *}(k_1)C_{\cal RS *}(k_2) + (\vec{k} {\rm \text { cyclic perms}})\right),
\end{aligned}
\end{equation}
and so on.

\subsection{Non-Gaussianity via power spectrum at horizon crossing}\label{subsec:SectionHorizon}

The non-Gaussianity parameter in its general form, as presented in \eqref{fNLGeneral}, is somewhat implicit. To gain further insight into the new contributions, let us analyze it in detail. Our goal is to express Eq. \eqref{fNLGeneral} in terms of background quantities, correlators at horizon crossing and transfer functions in order to obtain an analytical understanding of this equation and non-Gaussianity generation in rapid-turn attractor models.

If the expression for the number of e-folds $N(\phi^a)$ is known in terms of background quantities, one may explicitly insert the known expressions for $N_a$ and $N_{ab}$ into \eqref{fNLGeneral} and find the value of $f_{\rm NL}^{(4)}$. In the absence of such explicit expression, it is convenient to use the equations relating $N_a$ and $N_{ab}$ to the transfer function $T_\mathcal{RS}$ obtained in Eqs. \eqref{Na} and \eqref{Nab}, respectively. These can be used, along with Eqs. \eqref{nablaeN} and \eqref{theta} to find an expression for the non-Gaussianity parameter in Eq. \eqref{fNLGeneral}. We find the result to be rather cumbersome and full expressions are shown in Appendix ~\Ref{app:FullexpressionHorizon}, Eq. \eqref{FullNumeratorHorCrossing}. Here we show it in the schematic form 
\begin{equation}\label{numeratorHorizon}
\begin{aligned}
  N_a N_b N_{cd} K^{abcd}(k_1, k_2, k_3)&= a_1 \, {\cal P}_{{\cal R}_*}(k_1) {\cal P}_{{\cal R}_*}(k_2)+a_2 \, {\cal P}_{{\cal S}_*}(k_1) {\cal P}_{{\cal S}_*}(k_2)+a_3 \,  {\cal C}_{{\cal RS}_*}(k_1) {\cal C}_{{\cal RS}_*}(k_2)\\
  &+a_4 \, {\cal P}_{{\cal R}_*}(k_1) {\cal P}_{{\cal S}_*}(k_2)
   +
  a_5 \, {\cal P}_{{\cal S}_*}(k_1) {\cal P}_{{\cal R}_*}(k_2)+ a_6 \, {\cal C}_{{\cal RS}_*}(k_1) {\cal P}_{{\cal R}_*}(k_2)\\
  &+a_7 \,{\cal P}_{{\cal R}_*}(k_1) {\cal C}_{{\cal RS}_*}(k_2)+ a_8 \,{\cal C}_{{\cal RS}_*}(k_1) {\cal P}_{{\cal S}_*}(k_2)
   + a_9 \,{\cal P}_{{\cal S}_*}(k_1) {\cal C}_{{\cal RS}_*}(k_2)\\
    &+ (\vec{k} {\rm \text { cyclic perms}}),
\end{aligned}
\end{equation}
where coefficients $a_i$ are functions of background quantities and $T_{{\cal RS}}$. Similarly, one can compute the denominator of \eqref{fNLGeneral} to obtain
\begin{equation}\label{DenominatorHorizonCross}
\begin{aligned}
    N_e N_f N_g N_h K^{efgh}(k_1, k_2, k_3)&= \\
    \left( P_{{\cal R}_*}(k_1)+2 T_{{\cal RS}} C_{{\cal RS}_*}(k_1)+ T^2_{{\cal RS}} P_{{\cal S}_*}(k_1) \right)&
      \left( P_{{\cal R}_*}(k_2)+2 T_{{\cal RS}} C_{{\cal RS}_*}(k_2)+ T^2_{{\cal RS}} P_{{\cal S}_*}(k_2) \right)\\
      &+ (\vec{k} {\rm \text { cyclic perms}}).
\end{aligned}
\end{equation}
The result above can also be obtained using the denominator of the form \eqref{fNLdef} with $P_\mathcal{R}$ expressed via power spectrum at horizon crossing using \eqref{PRS_transfer}. Clearly, Eq. \eqref{DenominatorHorizonCross} can be written in the same form as \eqref{numeratorHorizon} but with some other coefficients $b_i$ instead of $a_i$, that however have explicit dependence only on the transfer function $T_{{\cal RS}}$. 
After dividing numerator by denominator we see that, in general, the resulting $f_{\rm{NL}}$ is scale dependent. 

To make further progress in analysis, we assume a scale-invariant power spectrum of perturbations at horizon crossing. This allows us to set  $P_{{\cal R}_*}(k)=P_{{\cal R}_*},\, P_{{\cal S}_*}(k)=P_{{\cal S}_*},\, C_{{\cal RS}_*}(k)=C_{{\cal RS}_*}$, however keeping all power spectra to be distinct from each other, i.e. $P_{{\cal R}_*}\neq P_{{\cal S}_*} \neq  C_{{\cal RS}_*}$. Note that $k_*$ dependence remains present via the horizon crossing time $t_*$ and the relation $k_*=a(t_*)H(t_*)$. It is worth pointing out that in the derivation of commonly used $\delta N$ formula \eqref{Ksimple} it is implicitly assumed that the horizon crossing time of wavenumbers $k_1, k_2, k_3$ is identified with a single time  $t_*$. This corresponds to a near-equilateral momentum regime with $k_1\approx k_2\approx k_3$. A mild hierarchy between
the scales $|\log k_1/k_3|\sim {\cal O}(1)$ was considered in \cite{Byrnes:2009pe, Dias:2013rla}, which is sometimes referred as the mild squeezing regime. In this case the scale dependence of $f_{\rm NL}$ was shown to be first order in slow-roll \cite{Byrnes:2009pe}. In order to account for a highly squeezed momentum configuration $k_1\ll  k_2 \approx k_3$ one needs to take into account very different horizon exit times $t_1\ll  t_2 \approx t_3$. In \cite{Kenton:2015lxa} this was discussed in detail and $\delta N$ expressions were extended in order to allow for multiple horizon crossing times. It was found that for some models there is a correction at a level of 20\% in the highly squeezed limit compared to expressions in the mildly squeezed limit. As a first step, in our computation we adopt the assumption of single horizon crossing  time $t_*$ and leave extensive investigation of scale and shape dependence for forthcoming work. 

It is convenient to introduce parameters that encode relative magnitudes of power spectra.
Let us define the ratios of isocurvature to curvature perturbations $\alpha{_*}$ and the cross-correlation ratio $\beta_{*}$ at horizon crossing as
\begin{equation}\label{alphabetahorizon}
    \alpha{_*} \equiv  \frac{{\cal P}_{{\cal S}_*}}{{\cal P}_{{\cal R}_*}}, \quad \beta_{*} \equiv \frac{{\cal C}_{{\cal RS}_*}}{\sqrt{{\cal P}_{{\cal R}_*}{\cal P}_{{\cal S}_*}}}.
\end{equation}
where the cross-correlation ratio is defined similarly as in \cite{Bartolo:2001rt}. 
Now, after dividing \eqref{FullNumeratorHorCrossing} by \eqref{DenominatorHorizonCross} with a scale-invariant assumption for power spectra and expressing it in terms of $\alpha{_*}$ and $\beta_{*}$ via \eqref{alphabetahorizon}, we obtain the following result
\begin{equation}\label{fNLHorizonEpsTheta}
\begin{aligned}
        -\frac{6}{5}  f^{(4)}_{\rm NL}=\frac{\sqrt{2\epsilon_*}}{(1+2\beta_* \sqrt{\alpha_*}\, T_{\cal RS}+\alpha_* T_{\cal RS}^2 )^2}\left[
        (\nabla T_{\cal RS})_{\parallel *}(\beta_* \sqrt{\alpha_*}\, +\alpha_* T_{\cal RS})(1+\beta_* \sqrt{\alpha_*}\,  T_{\cal RS}) \right. \\
         \left. \, +(\nabla T_{\cal RS})_{\perp *}(\beta_* \sqrt{\alpha_*}\, +\alpha_* T_{\cal RS})^2 \right. \\
       \left. \,  -\frac{(\nabla \epsilon)_{\parallel *}}{2\epsilon_*}(1+\beta_* \sqrt{\alpha_*}\, T_{\cal RS})(1+2\beta_* \sqrt{\alpha_*}\, T_{\cal RS}+\alpha_* T_{\cal RS}^2 ) \right. \\
     \left. \,  -\frac{(\nabla \epsilon)_{\perp *}}{2\epsilon_*}(\beta_* \sqrt{\alpha_*}\, +\alpha_* T_{\cal RS})(1+2\beta_* \sqrt{\alpha_*}\, T_{\cal RS}+\alpha_* T_{\cal RS}^2 ) \right. \\
      \biggl. \, - \left(-\beta_* \sqrt{\alpha_*}\, +T_{\cal RS}(1-\alpha_*+\beta_* \sqrt{\alpha_*}\, T_{\cal RS})\right)\left(\theta_{\parallel *}(1+\beta_* \sqrt{\alpha_*}\, T_{\cal RS})+\theta_{\perp *}(\beta_* \sqrt{\alpha_*}\, +\alpha_* T_{\cal RS})\right) \biggr].
 \end{aligned}
   \end{equation}
To make the final formula more intuitive, let us separate the power spectrum quantities from background ones and combine terms that are proportional to the turn-rate, speed up rate and components of background mass matrix. In order to do that, we use \eqref{epsParPerp} and \eqref{thetaParPerp} and obtain the final expression for $f_{\rm NL}$ in the form
\begin{equation}\label{fNLviaIhorizon}
\begin{aligned}
      -\frac{6}{5}f^{(4)}_{\rm NL}=\sqrt{2\epsilon_*}(\nabla T_{\cal RS})_{\parallel *}I_{1*} +
  \sqrt{2\epsilon_*}(\nabla T_{\cal RS})_{\perp *}I_{2*} 
  -\frac{\eta_{\parallel *}}{\sqrt{2\epsilon_*}}I_{3*} -\frac{\eta_{\perp *}}{\sqrt{2\epsilon}}I_{4*}\\
  +\left(\tilde{M}_{\perp\perp *}+  \frac{\sqrt{2\epsilon_*}}{(3-\epsilon_*)^2}\left(\eta_{\parallel *}\tilde{M}_{\perp\perp *}- \eta_{\perp *}\tilde{M}_{\perp\parallel *}  \right)\right)I_{5*}
  + \tilde{M}_{\perp\parallel *} I_{6*}.
   \end{aligned}
   \end{equation}
Please, note that the formula is exact and does not rely on the slow-roll approximation.  The coefficients $I_i$ depend on time through $T_{\cal RS}$ and are given by
\begin{equation}\label{Ihorizon}
\begin{aligned}
    I_{1*}&=\frac{(\beta_* \sqrt{\alpha_*}\, +\alpha_* T_{\cal RS})(1+\beta_* \sqrt{\alpha_*}\,  T_{\cal RS})}{(1+2\beta_* \sqrt{\alpha_*}\, T_{\cal RS}+\alpha_* T_{\cal RS}^2 )^2},\\
    I_{2*}&=\frac{(\beta_* \sqrt{\alpha_*}\, +\alpha_* T_{\cal RS})^2}{(1+2\beta_* \sqrt{\alpha_*}\, T_{\cal RS}+\alpha_* T_{\cal RS}^2 )^2},\\
    I_{3*}&=\frac{(1+\beta_* \sqrt{\alpha_*}\, T_{\cal RS} )}{(1+2\beta_* \sqrt{\alpha_*}\, T_{\cal RS}+\alpha_* T_{\cal RS}^2 )},\\
    I_{4*}&=\frac{(1+\beta_* \sqrt{\alpha_*}\, T_{\cal RS} )\left(-\beta_* \sqrt{\alpha_*}\, +T_{\cal RS}(1-\alpha_*+\beta_* \sqrt{\alpha_*}\, T_{\cal RS})\right)}{(1+2\beta_* \sqrt{\alpha_*}\, T_{\cal RS}+\alpha_* T_{\cal RS}^2 )^2},\\
    I_{5*}&=\frac{(\beta_* \sqrt{\alpha_*}\, +\alpha_* T_{\cal RS} )\left(-\beta_* \sqrt{\alpha_*}\, +T_{\cal RS}(1-\alpha_*+\beta_* \sqrt{\alpha_*}\, T_{\cal RS})\right)}{(1+2\beta_* \sqrt{\alpha_*}\, T_{\cal RS}+\alpha_* T_{\cal RS}^2 )^2},\\
    I_{6*}&=\frac{(\beta_* \sqrt{\alpha_*}\, +\alpha_* T_{\cal RS})}{(1+2\beta_* \sqrt{\alpha_*}\, T_{\cal RS}+\alpha_* T_{\cal RS}^2 )}.
\end{aligned}
\end{equation}

\begin{figure}
\includegraphics[width=.48\columnwidth]{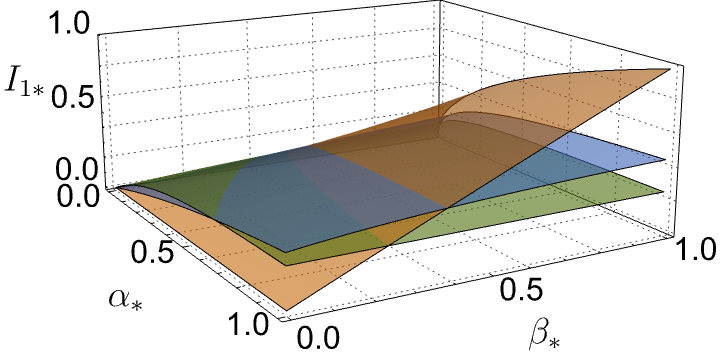} \hspace{0.4 cm}
\includegraphics[width=.48\columnwidth]{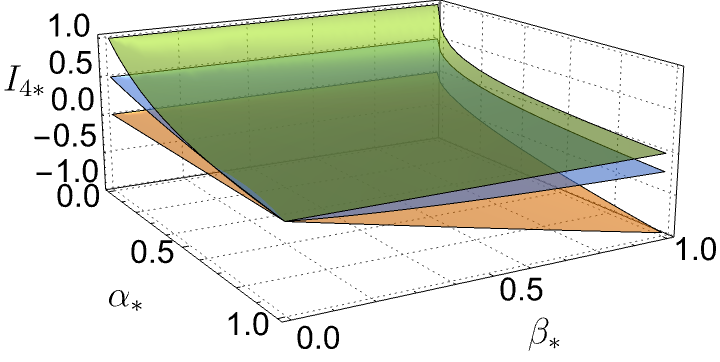}\\\vspace{0.4 cm}
\includegraphics[width=.48\columnwidth]{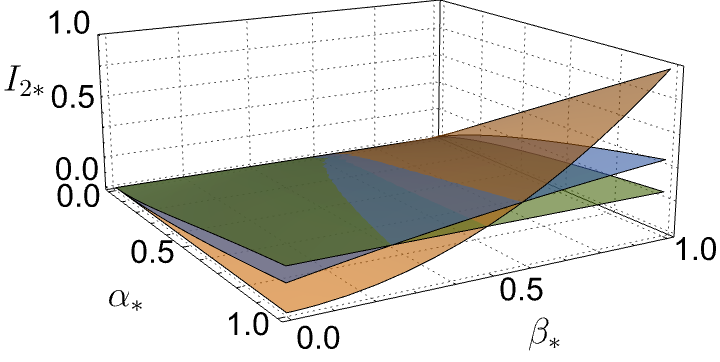}\hspace{0.5 cm}
\includegraphics[width=.48\columnwidth]{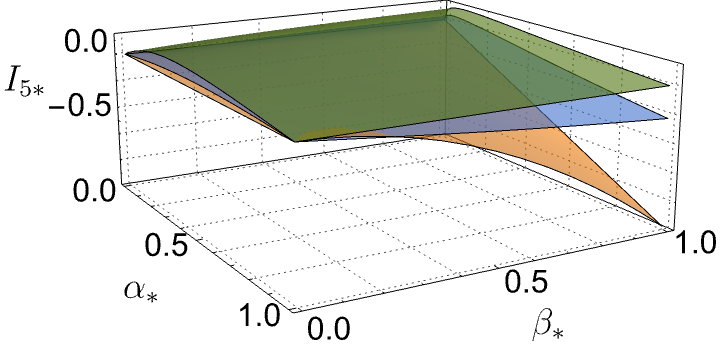}\\ \vspace{0.4 cm}
\includegraphics[width=.48\columnwidth]{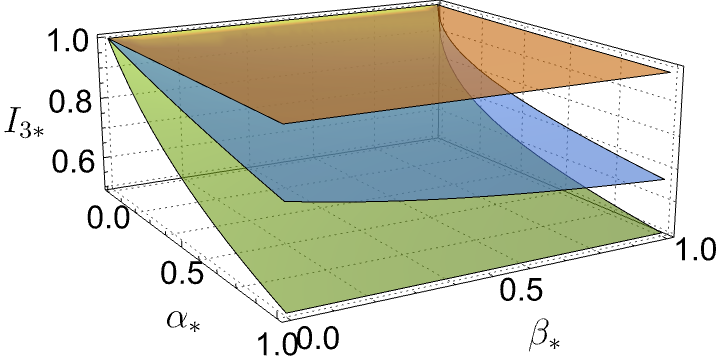}\hspace{0.5 cm}
\includegraphics[width=.48\columnwidth]{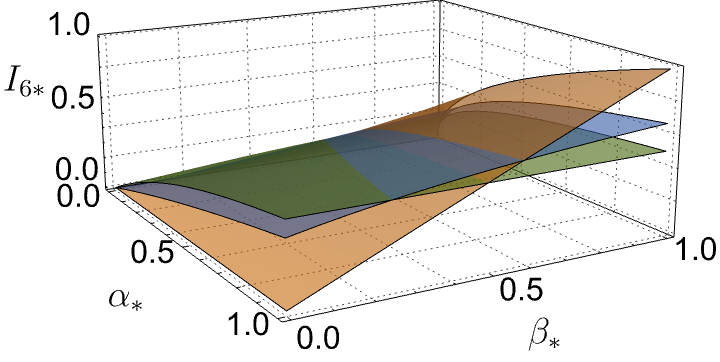}
\caption[]{  3D plots for the coefficients $I_{i*}$ at horizon crossing and their dependence on $\alpha_*=P_{\cal S*}/P_{\cal R*}$ and $\beta_*=C_{\cal RS*}/\sqrt{P_{\cal R*} P_{\cal S*}}$ for $T_{\cal RS}=0.01\, , 0.5\, , 1$ (brown, blue and green surfaces respectively).}\label{fig:Ihor}
\end{figure}
On Figure \ref{fig:Ihor} we show the dependence of coefficients $I_i$ on $\alpha_*$ and $\beta_*$ for different values of the transfer function $T_{\cal RS}$.  All  coefficients $I_{i*}$ are bounded to be less than unity, except of the coefficient $I_{4*}$. For larger values of $T_{\cal RS}$, small $\alpha_*$ and any value of $\beta_*$ it  becomes larger than unity and can provide a significant contribution to \eqref{fNLviaIhorizon}.

\subsection{Comparison with SRST approximation}\label{subsec:ComparisonSRST}
Let us now compare our result of the non-Gaussianity parameter with the one obtained in SRST approximation in \cite{Peterson:2010mv}. In order to do that, we set ${\cal C}_{{\cal RS}_*}=0,\, {\cal P}_{{\cal R}_*}={\cal P}_{{\cal S}_*}$ as in \cite{Peterson:2010mv}. This implies $\alpha{_*}=1$ and $\beta_{*}=0$ and we find the coefficients \eqref{Ihorizon} reduce to
\begin{alignat}{3}
  I_{1*} &= \frac{T_{\cal RS}}{(1+T_{\cal RS}^2)^2}, \quad I_{3*} &=\frac{1}{(1+T_{\cal RS}^2)},   \quad I_{4*}&=0,\\
  I_{2*} &=\frac{T_{\cal RS}^2}{(1+T_{\cal RS}^2)^2},  \quad I_{6*} &=\frac{T_{\cal RS}}{(1+T_{\cal RS}^2)},   \quad I_{5*}&=0.
\end{alignat}
Next, we plug them into \eqref{fNLviaIhorizon} and get
\begin{align}
\begin{split}
    -\frac{6}{5}  f_{\rm NL}^{(4),\, \rm SRST} =~ \sqrt{2\epsilon_*}\biggr[&\cos \Delta_N^2\left(-\frac{(\nabla \epsilon)_{\parallel *}}{2\epsilon_*}+\sin \Delta_N\cos \Delta_N (\nabla T_{\cal RS})_{\parallel *}\right)\\&+\sin \Delta_N\cos \Delta_N\left(-\frac{(\nabla \epsilon)_{\perp *}}{2\epsilon_*}+\sin \Delta_N\cos \Delta_N (\nabla T_{\cal RS})_{\perp *}\right) \biggr], \label{fNLslst}
\end{split}
\end{align}
where we used equations \eqref{epsParPerp}, \eqref{MtildeViaEta} and the definition of the correlation angle \eqref{correlationAngle} to re-express $ \cos \Delta_N=\frac{1}{\sqrt{1+T_{\cal RS}^2}}$ and $ \sin \Delta_N=\frac{T_{\cal RS}}{\sqrt{1+T_{\cal RS}^2}}$. We see that this result is exactly of the same form as the one obtained in \cite{Peterson:2010mv}. An implicit difference is, however, contained in the term $(\nabla \epsilon)_{\perp *}/\sqrt{2\epsilon_*}=-\tilde{M}_{\perp\parallel *}$. In the case of the SRST approximation, one has $\tilde{M}_{\perp\parallel *}=M_{\perp\parallel *}=M_{\parallel\perp *}\approx -\eta_{\perp *}/\sqrt{2\epsilon_*}\ll 1$, making this contribution negligible if one looks for $\mathcal{O}(1)$ non-Gaussianity. On the other hand, for rapid-turn models, the matrix component $\tilde{M}_{\perp\parallel *}$ can be significant and potentially generate large non-Gaussianity. As pointed out in \cite{Peterson:2010mv}, the term containing $(\nabla T_{\cal RS})_{\perp *}$ can also lead to large non-Gaussianities, in particular when the sourcing of curvature perturbations from isocurvature perturbations is highly sensitive to initial conditions of the inflationary trajectory. This term can be expressed via background quantities and transfer functions in some analytically solvable models as was shown in \cite{Peterson:2010mv}.

In the single-field (SF) limit $T_{\cal RS}=0$ and $M_{\perp\parallel *}=0$ and the second parenthesis in \eqref{fNLslst} is identically zero. Only the first term with quantities that are parallel to the field trajectory is non-zero.  In \cite{Peterson:2010mv} it was shown that it is equal to
\begin{equation}
  -\frac{6}{5}  f_{\rm NL}^{(4),\,\rm SF} =  \frac{1}{2}\left(n_s-1-n_t\right),
\end{equation}
where $n_s$ is the scalar spectral index and $n_t$ is the tensor tilt.
The single-field consistency relation has contributions from both $f_{\rm NL}^{(3),\,\rm SF }$ and $f_{\rm NL}^{(4),\,\rm SF}$. Taking into account $f_{\rm NL}^{(3),\,\rm SF}$, one gets \cite{Peterson:2011yt}
\begin{equation}
  -\frac{6}{5}  f_{\rm NL}^{\rm SF} =  \frac{1}{2}\left(n_s-1\right).
\end{equation}
which coincides with the Maldacena single-field consistency condition \cite{Maldacena:2002vr}.

\subsection{Non-Gaussianity via power spectrum after horizon crossing}\label{subsec:SectionEOI}

In this section we will follow the same steps as in Section \ref{subsec:SectionHorizon} in order to obtain the non-Gaussianity parameter via the power spectra evaluated at the end of inflation or any other time after horizon crossing. Specifically,  we express the power spectra at horizon crossing via the power spectra at later time using \eqref{PRS_transfer} and, together with $N_a$ and $N_{ab}$ from Eqs. \eqref{Na} and \eqref{Nab}, insert into the general formula \eqref{fNLGeneral}.
The resulting numerator is of the form 
\begin{gather}\label{numeratorEnd}
    N_a N_b N_{cd} K^{abcd}(k_1, k_2, k_3)=\\
 c_1\,P_{\cal{R}}(k_1)P_{\cal{R}}(k_2)+  c_2\,C_{\cal{RS}}(k_1)C_{\cal{RS}}(k_2)+c_3\,P_{\cal{R}}(k_1)C_{\cal{RS}}(k_2)+c_4\,C_{\cal{RS}}(k_1)P_{\cal{R}}(k_2)+ (\vec{k} {\rm \text { cyclic perms}}), \notag
\end{gather}
where coefficients $c_i$ are functions of background quantities, $T_{{\cal RS}}$ and $T_{\cal SS}$. One can see that in \eqref{numeratorEnd} there is no explicit dependence on $P_{\cal{S}}(k)$, however the presence of isocurvature perturbations is implicitly encoded via $T_{\cal SS}$ as we will see below. The denominator at the end of inflation is given by the definition \eqref{fNLdef} as the permutation 
\begin{equation}\label{denominatorEnd}
  N_e N_f N_g N_h K^{efgh}(k_1, k_2, k_3)=   P_{\cal R}(k_1)P_{\cal R}(k_2)+ (\vec{k} {\rm \text { cyclic perms}}).
\end{equation}
 Therefore, taking the ratio of \eqref{numeratorEnd} to \eqref{denominatorEnd}, we find that an overall structure of the non-Gaussianity parameter is of the form
\begin{equation}\label{fNLRC}
     -\frac{6}{5}   f_{\rm NL}^{(4)}(k_1,k_2,k_3)=\sum_{IJ} f^{IJ}_{\rm NL}\frac{\tilde{{\cal P}}^I(k_1)\tilde{{\cal P}}^J(k_2)+(\vec{k} {\rm \text { cyclic perms}})}{P_{\cal R}(k_1)P_{\cal R}(k_2)+ (\vec{k} {\rm \text { cyclic perms}})},
\end{equation}
with $I,J= {\cal R, C}$ and $\tilde{{\cal P}}^{{\cal R}}(k)=P_\mathcal{R}(k)$, $\tilde{{\cal P}}^{{\cal C}}(k)=C_\mathcal{RS}(k)$.
From the full expression given in equation \eqref{FullNumeratorEnd} in Appendix, we find the components $f^{IJ}_{\rm NL}$ explicitly as
 \begin{align}
     f^{{\cal R}{\cal R}}_{\rm NL}&=\sqrt{2\epsilon_*}\left[-\frac{(\nabla \epsilon)_{\parallel *}}{2\epsilon_*}-\theta_{\parallel *}T_{{\cal RS}}
 \right],\\
   f^{{\cal C}{\cal C}}_{\rm NL}&=\frac{\sqrt{2\epsilon_*}}{T_{\cal SS}^2}\left[(1+T_{{\cal RS}}^2)(\theta_{\perp *}-T_{{\cal RS}}\theta_{\parallel *} )+((\nabla T_{\cal RS})_{\perp *}-T_{{\cal RS}}(\nabla T_{\cal RS})_{\parallel *} )
 \right],\\
 f^{{\cal R}{\cal C}}_{\rm NL}&=\frac{\sqrt{2\epsilon_*}}{T_{\cal SS}}\left[\theta_{\parallel *}(1+T_{{\cal RS}}^2)+(\nabla T_{\cal RS})_{\parallel *}
 \right],\\
  f^{{\cal C}{\cal R}}_{\rm NL}&=\frac{\sqrt{2\epsilon_*}}{T_{\cal SS}}\left[-\frac{(\nabla \epsilon)_{\perp *}-(\nabla \epsilon)_{\parallel *} T_{{\cal RS}} }{2\epsilon}-T_{{\cal RS}}(\theta_{\perp *}-\theta_{\parallel *} T_{{\cal RS}})
 \right].
 \end{align}
 One can see that when $\tilde{{\cal P}}^I(k)$ and $\tilde{{\cal P}}^J(k)$ have a distinct scale dependence, the resulting non-Gaussianity parameter in Eq. \eqref{fNLRC} also acquires a scale dependence. In addition, this parameter also depends on the shape functions of the $ {\cal CC},\,{\cal CR},\, {\cal RC}$ contributions. Here only the contribution $f^{{\cal R}{\cal R}}_{\rm NL}$ is local in the sense described in Section \ref{subsec:bispectrum}. Setting the cross-correlation to zero eliminates the scale and shape dependence of the non-Gaussianity parameter and reduces it to the one obtained in SRST approximation.
 
\begin{figure}
\includegraphics[width=.48\columnwidth]{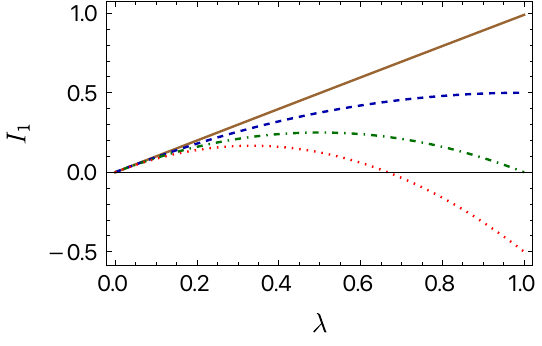}\hspace{0.5 cm}
\includegraphics[width=.48\columnwidth]{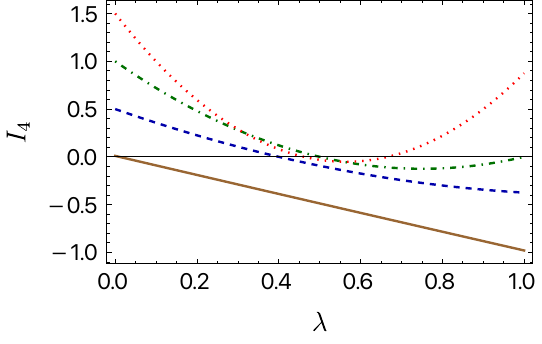}\\  \vspace{0.4 cm}
\includegraphics[width=.48\columnwidth]{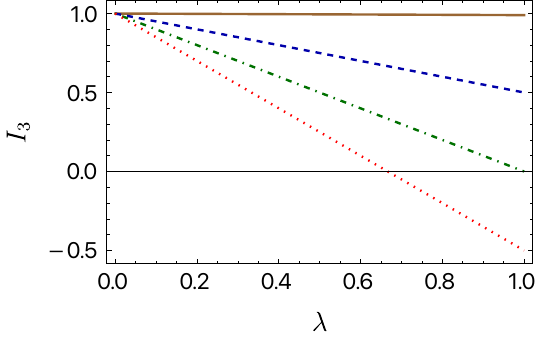}\hspace{0.5 cm}
\includegraphics[width=.48\columnwidth]{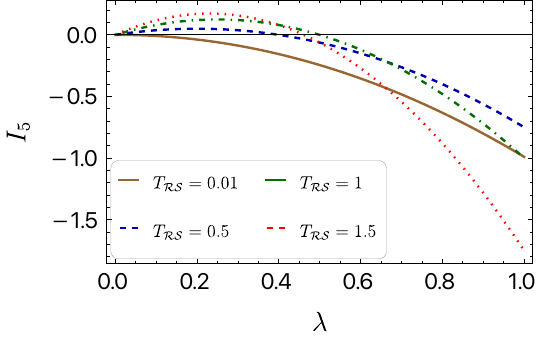}
\caption[]{ Coefficients $I_{1},\, I_{3}, \,I_{4},\, I_{5}$ and their dependence on the parameter $\lambda$ for different values of the transfer function at the end of inflation $T_{\cal RS}=0.01,\, 0.5,\, 1,\, 1.5$ (brown solid, blue dashed, green dot-dashed and red dotted respectively). Coefficients $I_{2}$ and $I_{6}$ do not depend on $T_{\cal RS}$ and grow quadratically and linearly (respectively) with $\lambda$, hence we omit those plots here.}\label{fig:Iend}
\end{figure}

 To proceed further, as before, we assume scale-invariant power spectrum of perturbations after horizon crossing, i.e. taking $P_{{\cal R}}(k)=P_{{\cal R}},\, P_{{\cal S}}(k)=P_{{\cal S}},\, C_{{\cal RS}}(k)=C_{{\cal RS}}$, with $P_{{\cal R}}\neq P_{{\cal S}} \neq  C_{{\cal RS}}$. In this case we find that the resulting non-Gaussianity parameter at any moment of time after horizon crossing may be parameterized similarly as in Section \ref{subsec:SectionHorizon}
\begin{equation}\label{fNL4end}
\begin{aligned}
      -\frac{6}{5}f^{(4)}_{\rm NL}=\sqrt{2\epsilon_*}(\nabla T_{\cal RS})_{\parallel *}I_1 +
  \sqrt{2\epsilon_*}(\nabla T_{\cal RS})_{\perp *}I_2 
  -\frac{\eta_{\parallel *}}{\sqrt{2\epsilon_*}}I_3 -\frac{\eta_{\perp *}}{\sqrt{2\epsilon}}I_4\\
  +\left(\tilde{M}_{\perp\perp*}+  \frac{\sqrt{2\epsilon_*}}{(3-\epsilon_*)^2}\left(\eta_{\parallel *}\tilde{M}_{\perp\perp *}- \eta_{\perp *}\tilde{M}_{\perp\parallel *}  \right)\right)I_5
  + \tilde{M}_{\perp\parallel *} I_6,
   \end{aligned}
   \end{equation}
where the coefficients $I_i$ depend on time via the transfer function $T_{\cal RS}$ and the parameter $\lambda$, and can be written in the form
\begin{equation}\label{Iend}
\begin{aligned}
I_1&=  \lambda\left(1-\lambda \,T_{\cal RS} \right),\\
    I_2&=\lambda^2,\\
    I_3&=\left(1-\lambda \, T_{\cal RS} \right),\\
    I_4&=\left(\lambda\, T_{\cal RS}-1\right)
    \left( -T_{\cal RS}  +\lambda(1+T_{\cal RS}^2)\right),\\
    I_5&=\lambda\left(T_{\cal RS} -\lambda\left(1+T_{\cal RS}^2\right)\right),\\
     I_6&=\lambda,
\end{aligned}
\end{equation}
with
\begin{equation}
\lambda\equiv \frac{{\cal C}_{{\cal RS}}}{\sqrt{{\cal P}_{{\cal R}}{\cal P}_{{\cal S}}}}\sqrt{\frac{{\cal P}_{{\cal S_*}}}{{\cal P}_{{\cal R}}}}. \label{tildeBeta}
\end{equation}

We show the dependence of $I_i$ coefficients on the parameter $\lambda$ on Figure \ref{fig:Iend}. One can see that for vanishing cross-correlation the coefficients $I_1, I_2, I_5, I_6$ are zero for all values of the transfer function $T_{\cal RS}$. Interestingly, contributions from $I_3$ and $I_4$ may be non-zero even for zero cross correlation. It is worth noting that the coefficients \eqref{Iend} coincide with the coefficients \eqref{Ihorizon} when $ P_\mathcal{R} , C_\mathcal{RS} , P_\mathcal{S} $ are re-expressed via $ P_\mathcal{R}^* , C_\mathcal{RS}^* , P_\mathcal{S}^* $ using \eqref{PRS_transfer}, and vice versa. In general, relation \eqref{fNL4end} gives the value of the non-Gaussianity parameter $f^{(4)}_{\rm NL}$ at any moment of time after horizon crossing.

\section{Example}\label{sec:Example}

In this section we are going to demonstrate how our general formula works in one example of a rapid-turn model of inflation. To show the full power of our result we consider a model with non-zero cross-correlation power spectrum either 
at horizon crossing or at the end of inflation as well as with a sustained rapidly-turning trajectory. `Angular inflation', introduced in \cite{Christodoulidis:2018qdw}, provides a simple example. It describes a dynamical attractor along the boundary of the Poincare disc
and supports a sustainable rapid-turn regime of inflation as we show below. Angular inflation has an action of the form \eqref{action} with the field-space metric given by
\begin{equation}
    G_{ab}=\frac{6\tilde{\alpha}}{\left( 1-\phi^2-\chi^2\right)^2}\delta_{ab},
\end{equation}
where $\tilde{\alpha}$ is the curvature parameter\footnote{In the original paper \cite{Christodoulidis:2018qdw} for $\tilde{\alpha}$ is used letter $\alpha$. We use tilde to avoid confusion with the ratio of isocurvature to curvature power spectrum which we define with $\alpha$.}. The potential is given by
\begin{equation}
    V(\phi,\chi)=\frac{\tilde{\alpha}}{2}\left(m_{\phi}^2\phi^2+m_{\chi}^2\chi^2\right).
\end{equation}
In a flat limit $\tilde{\alpha} \rightarrow \infty$ recovers usual quadratic potential when fields are properly rescaled. In terms of polar coordinates, the fields may be written as $\phi=r\cos{(\theta)}$ and $\chi=r\sin{(\theta)}$. In the `radial phase' the slow-roll approximation holds for both the radial field, $r$, and the angular field, $\theta$. However, for large hyperbolic curvature there is a region before the end of inflation where the slow-roll approximation does not hold anymore. As a result, the fields speed up and enter into a regime of angular inflation. In the angular phase the field trajectory proceeds along an angular direction, while the radial field, to a first approximation, is frozen.

\begin{figure}
\centering
\includegraphics[width=.51\columnwidth]{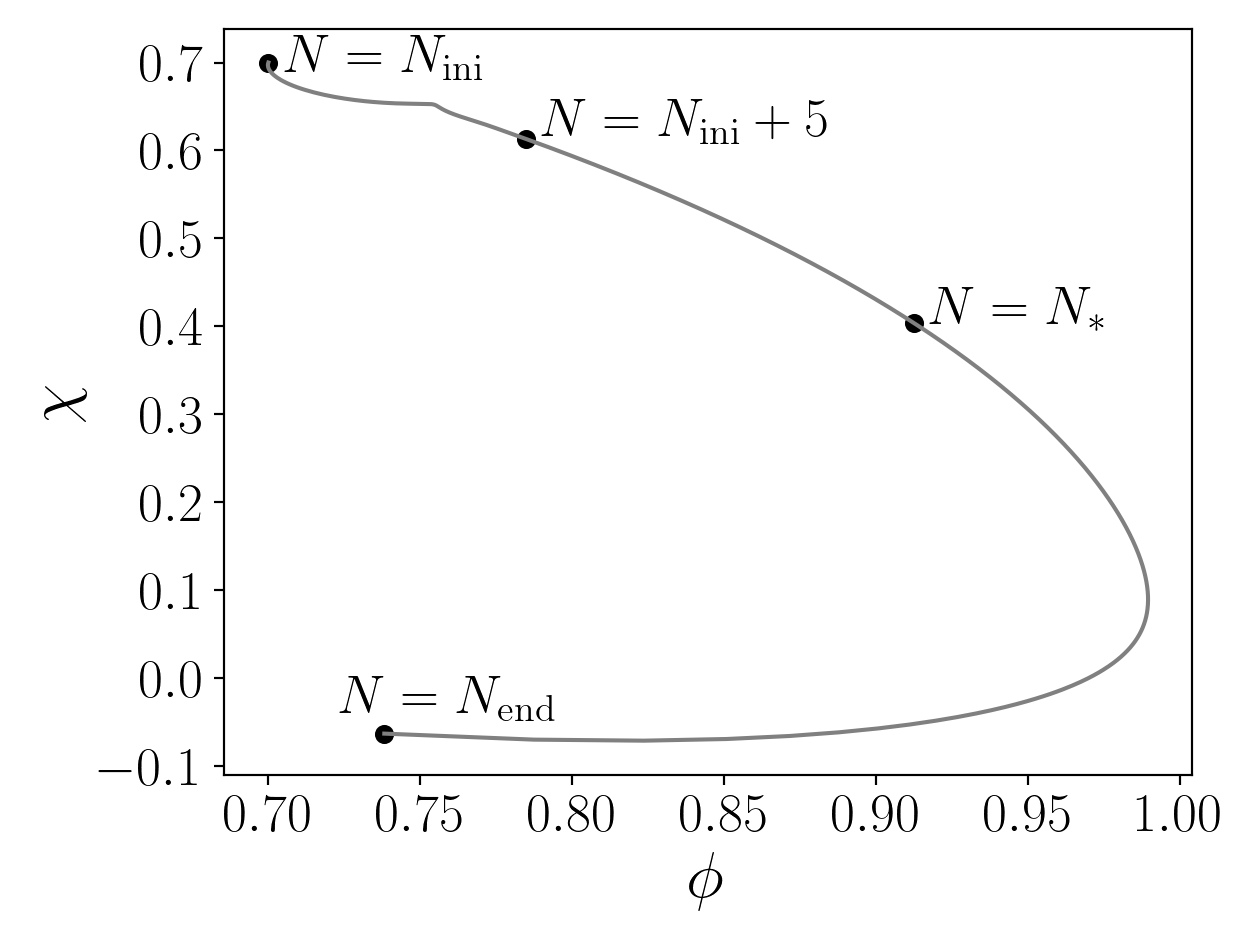}
\caption[]{Background trajectory of $\phi$ and $\chi$ fields. Black circles illustrate the number of e-folds $N$ at different points on the trajectory. The moment of horizon crossing corresponds to $N_*=N_{\rm end}-55$ and the end of inflation for this trajectory is at $N_{\rm end}=91.6$. 
}
\label{fig:background}
\end{figure}

We chose the following parameters 
\begin{equation}\label{params}
   \tilde{\alpha}=1/600, \quad R_m\equiv m_{\chi}^2/m_{\phi}^2=9,
\end{equation}
and tune initial conditions in such way to have a very short duration of the radial phase (a couple of first e-folds of inflation) and angular inflation phase throughout the rest of evolution. Background evolution for our choice of parameters is shown in Fig. \ref{fig:background}.

\begin{figure}
\includegraphics[width=.49\columnwidth]{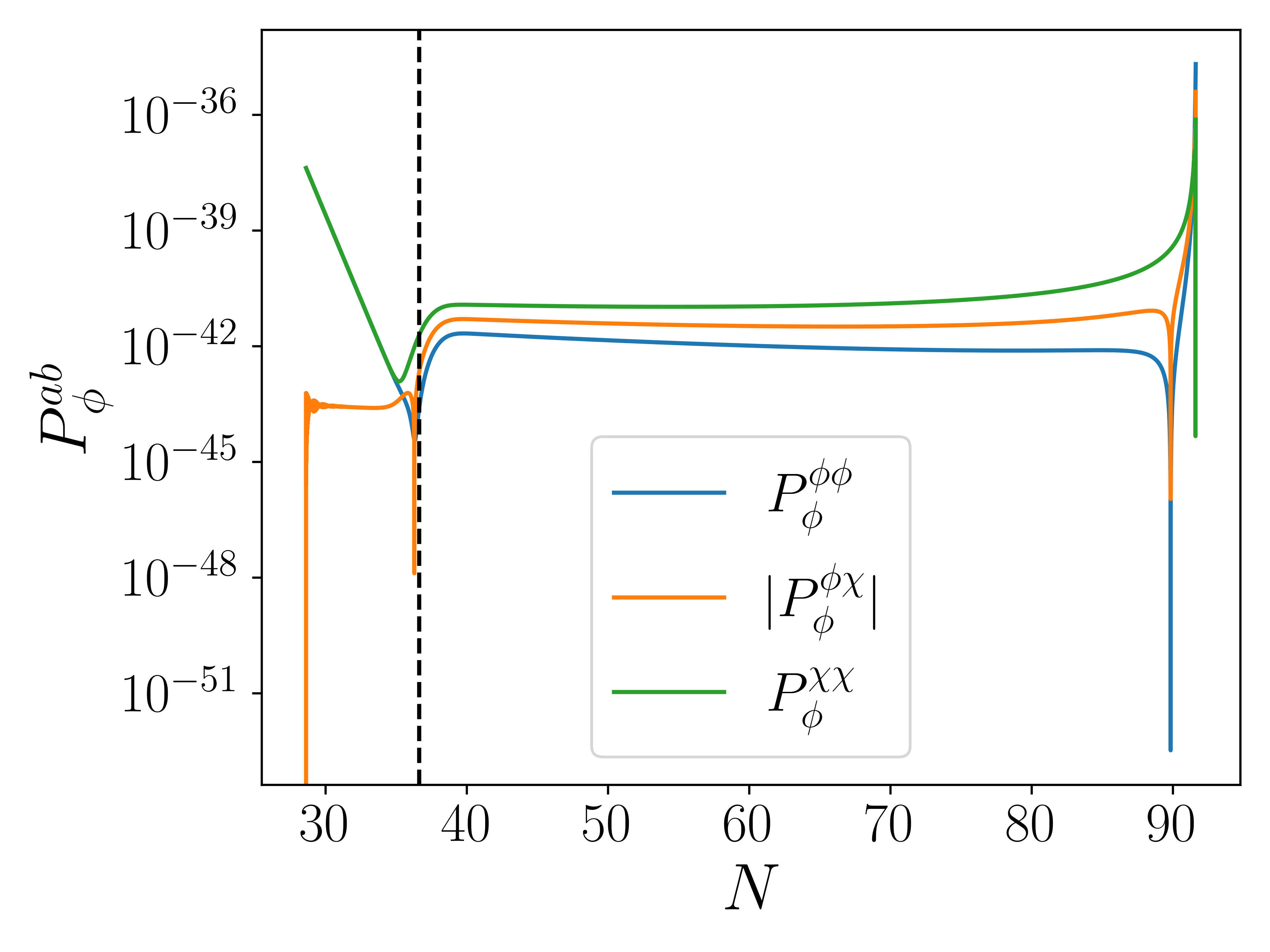}
\includegraphics[width=.49\columnwidth]{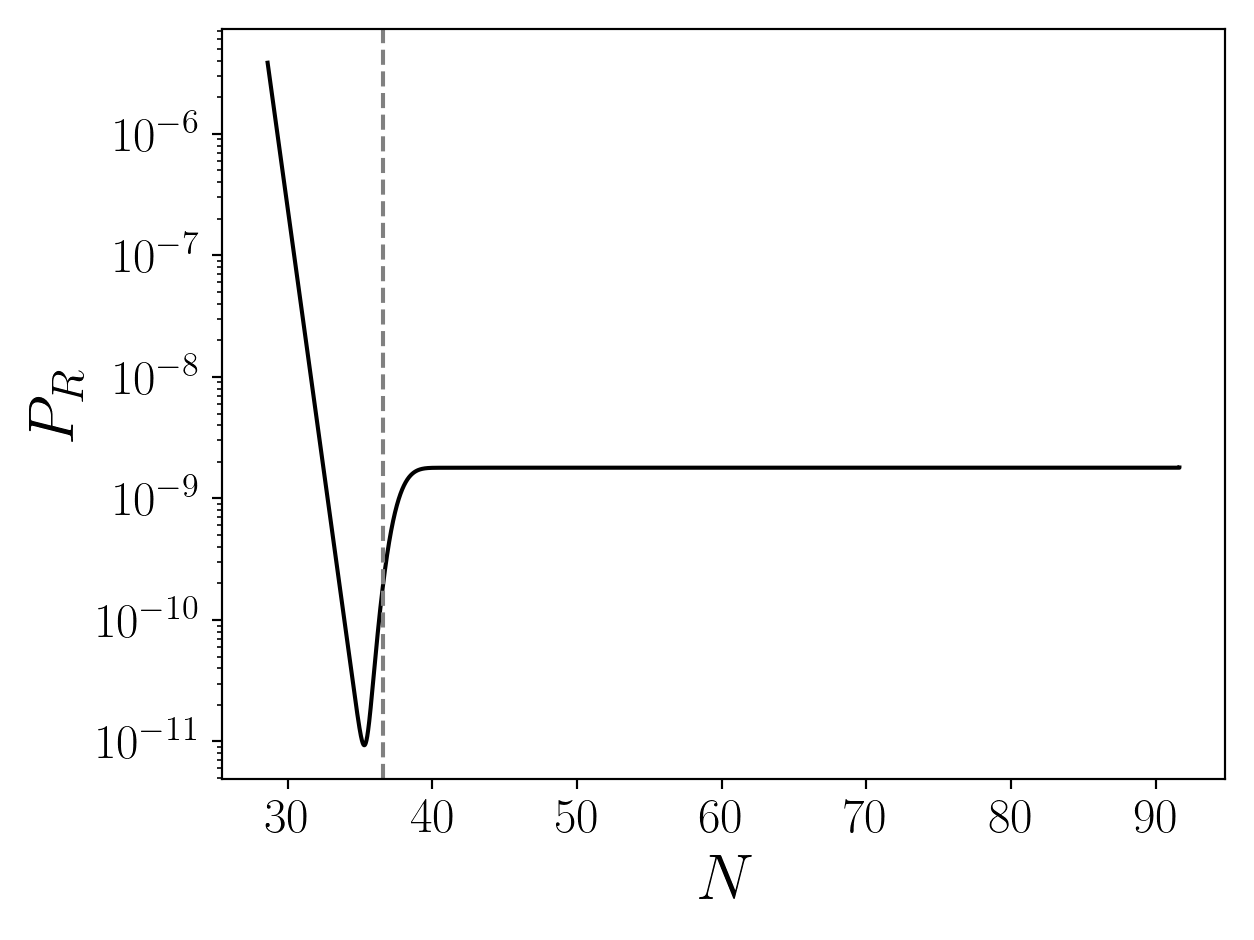}\\
\includegraphics[width=.49\columnwidth]{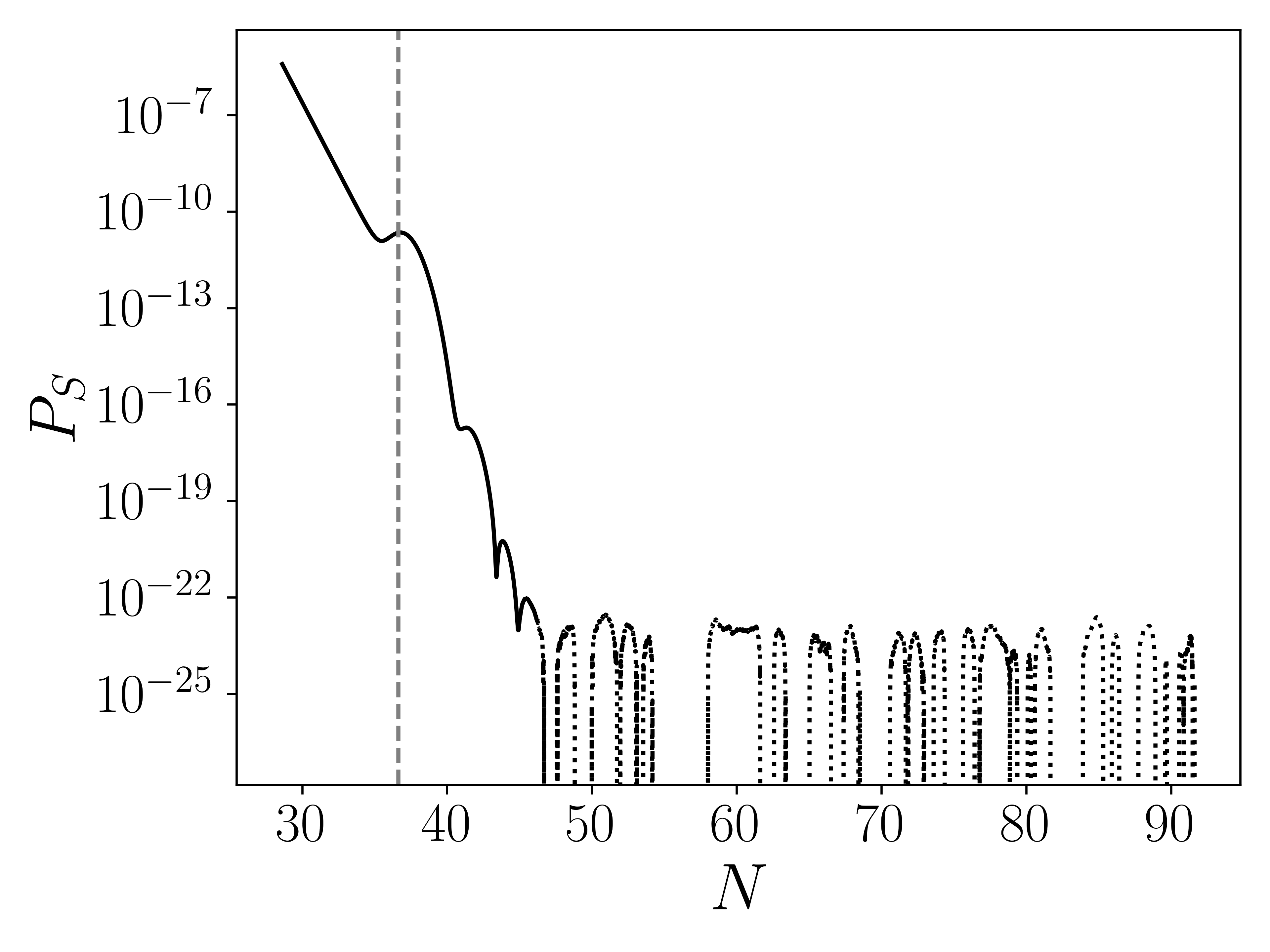}
\includegraphics[width=.49\columnwidth]{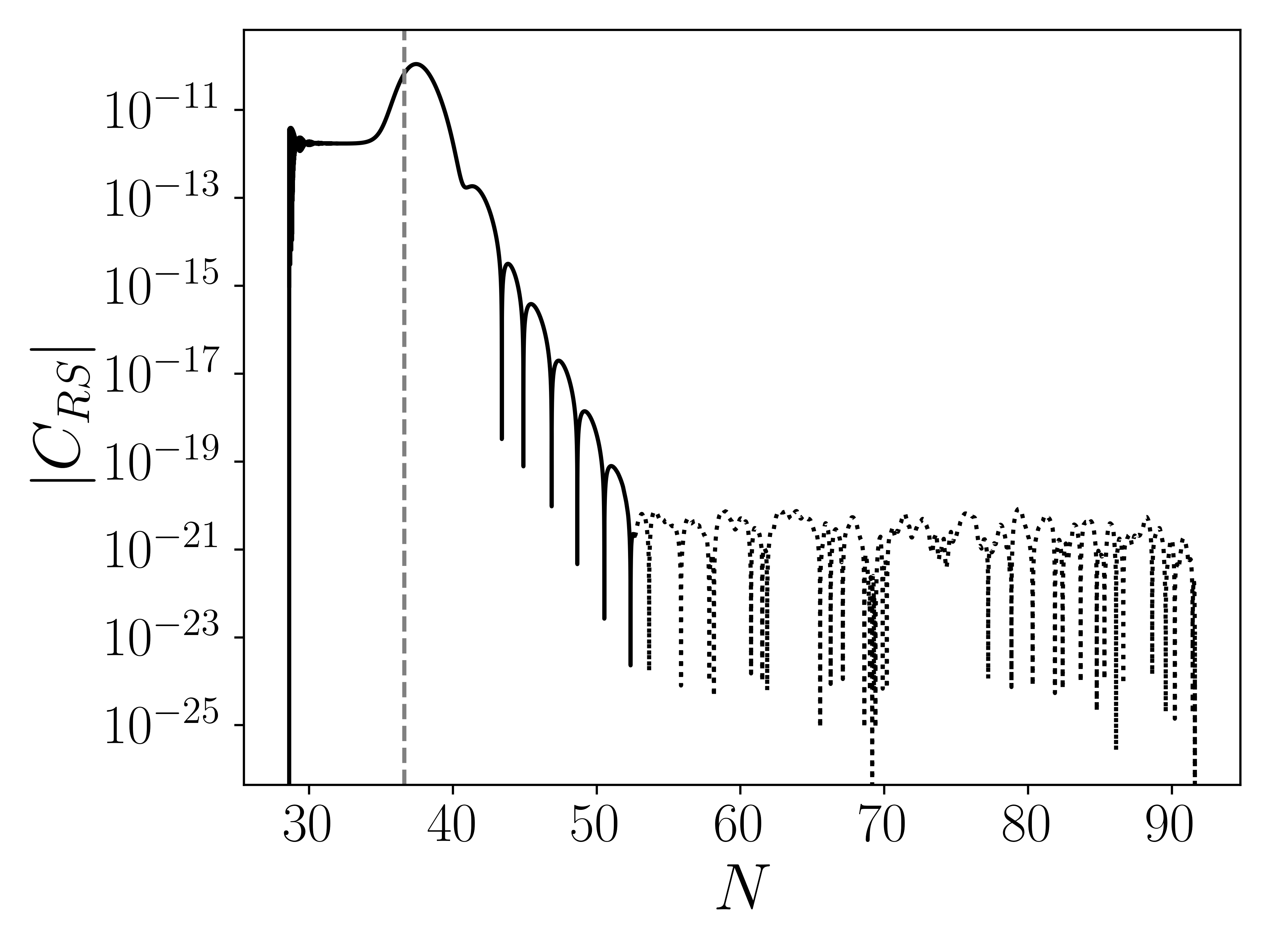}
\caption[]{Power spectrum of field perturbations $P^{ab}_{\phi}$, curvature perturbations $P_{\cal R}$, cross-correlation $|C_{\cal RS}|$ and isocurvature perturbations $P_{\cal S}$ throughout evolution (clockwise). Vertical dashed line on the plots corresponds to the moment of horizon crossing of the reference mode with a wave number $k_*$. Dotted lines on $|C_{\cal RS}|$ and $P_{\cal S}$ plots correspond to the region where the power spectra is numerically intractable. We believe in this region all of the isocurvature modes decay
away completely. We do not take into account numerical values from those regions in our computation.
}\label{fig:Power}
\end{figure}

To find the power spectra of perturbations we use the PyTransport code \cite{Mulryne:2016mzv}. The evolution of power spectrum \footnote{The value of the spectral tilt computed with the PyTransport code is $n_s=0.9652$.} with respect to the number of e-folds is shown in Fig. \ref{fig:Power}. For this parameter range the power spectrum of isocurvature perturbations, $P_{\cal S}$, together with the cross-correlation, $C_{\cal RS}$, decays rapidly and becomes negligible and numerically intractable. Therefore, due to poorly known correlation functions at the end of inflation, we will proceed with the expression \eqref{fNLviaIhorizon} for $f_{\rm NL}$ written via quantities computed at horizon crossing, where our numerical results are reliable.

\begin{figure}
    \centering
    \includegraphics[width=.44\columnwidth]{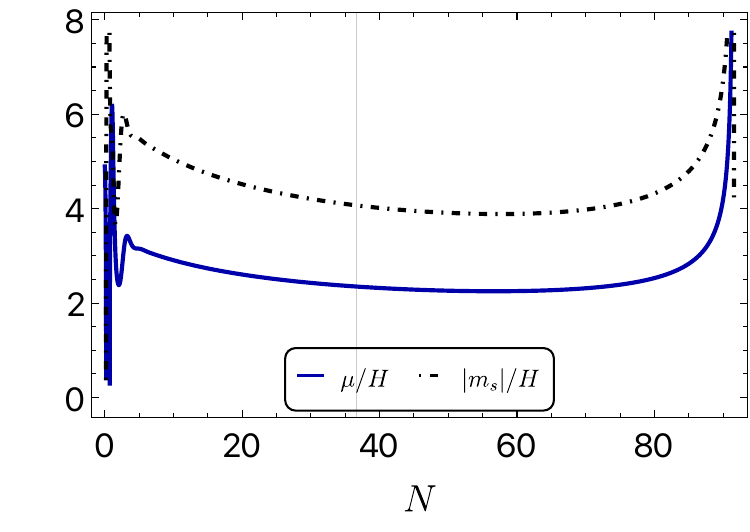}
    \includegraphics[width=.47\columnwidth]{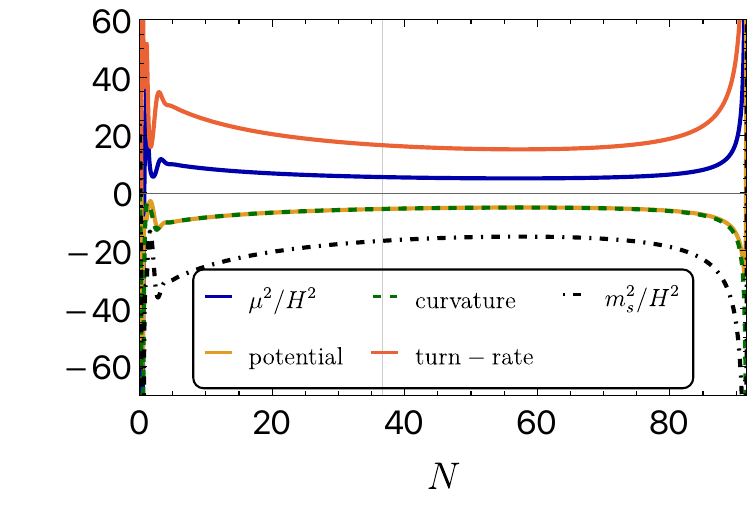}
    \caption{
    \textit{Left panel}: Evolution of entropic mass $\mu/H$ (blue solid curve) and the absolute value of $|m_s|/H$ (black dot-dashed curve) with respect to the number of e-folds. \textit{Right panel}: Different components of the entropic mass $\mu^2/H^2$ (solid blue curve): $e_{\perp}^a e_{\perp}^b\nabla_a\nabla_b V/H^2$ (solid orange curve), $-  e_{a, \perp} e_{\perp}^bR^a_{d f b}\phi'^d\phi'^f$ (green dashed curve), $3\left( \frac{\eta_{\perp}}{\sqrt{2\epsilon}}\right)^2$ (solid red curve). The black dot-dashed curve shows the evolution of $m_s^2/H^2$. The vertical gray line on both panels corresponds to the moment of horizon crossing.}
    \label{fig:mu}
\end{figure}

 We show the evolution of the entropic mass, defined in \eqref{mu}, and its contributions in Figure~\ref{fig:mu}. From the right panel of Fig.~\ref{fig:mu} we see that the contribution ${\cal M}_{\perp\perp}$, which is a sum of potential and curvature terms, is negative. However, since the turn rate is significant, the resulting mass of entropic fluctuation is of order one $\mu/H\sim {\cal O}(1)$, which is shown on the left panel of Fig.~\ref{fig:mu}, together with $|m_s|/H\sim {\cal O}(1)$ defined by
\begin{equation}
    m^2_s=\mu^2-4 H^2\left( \frac{\eta_{\perp}}{\sqrt{2\epsilon}}\right)^2={\cal M}_{\perp\perp}- H^2\left( \frac{\eta_{\perp}}{\sqrt{2\epsilon}}\right)^2.
\end{equation}
This explains the quick decay of isocurvature power spectrum shown in Fig.~\ref{fig:Power}. 
 Nevertheless, it is essential to note that the dynamics never reach the regime of $\mu/H\gg 1$ and cannot be fully described by the effective single-field approaches and regimes discussed in e.g.~\cite{Garcia-Saenz:2018ifx}. Still, flattened and equilateral contributions to the bispectrum are expected~\cite{Garcia-Saenz:2018ifx}, in addition to the contribution of the local shape sourced by the (decaying but) non-vanishing entropic modes on superhorizon scales. We will get back to the discussion of the resulting bispectrum shape for the current model later in this section. 

\begin{figure}
\includegraphics[width=.48\columnwidth]{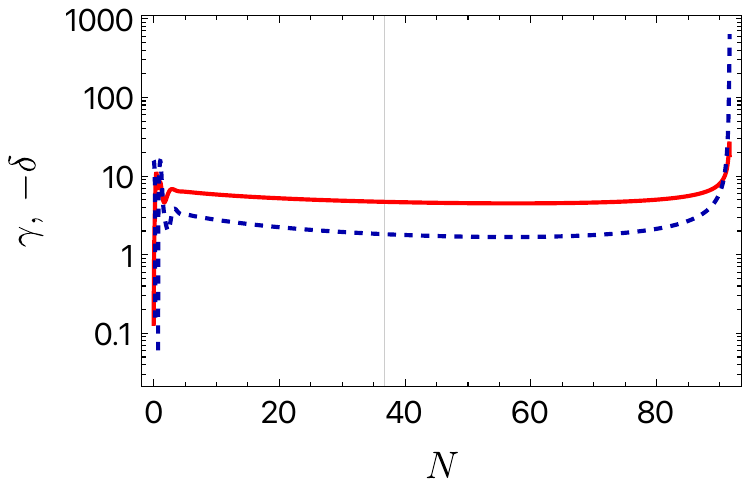}\hspace{0.4 cm}
\includegraphics[width=.47\columnwidth]{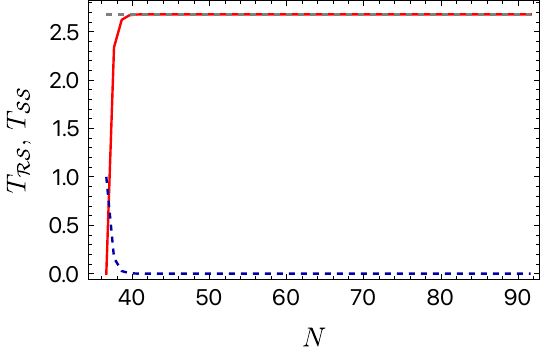}
\caption[]{{\it Left panel:} Evolution of functions $\gamma(N)$ (red solid) and $-\delta(N)$ (blue dashed) with respect to the number of e-folds. The vertical gray line corresponds to the moment of horizon crossing. {\it Right panel:} Transfer functions $T_{\cal{R S}}(N_*,N)$ (red solid) and $T_{\cal{S S}}(N_*,N)$ (blue dashed) and their evolution in e-folds. The horizontal gray dashed line shows
that soon after horizon crossing $T_{\cal{R S}}(N_*,N)$ saturates and does not get sourced anymore. In both panels we used parameters given in Eq. \eqref{params}. }
\label{fig:TRS}
\end{figure}

To compute the transfer function we use equations \eqref{TRSdef}, \eqref{gammaDeltaDef}, \eqref{MTRS}. In Fig. \ref{fig:TRS} we show the evolution of $\gamma(N)$, $-\delta(N)$, $T_{\cal{R S}}(N_*,N)$ and $T_{\cal{S S}}(N_*,N)$ during inflation. One can see that during angular phase of inflation $\delta(N)<0$, therefore isocurvature modes exponentially decay to zero. Nevertheless, positive and non-zero $\gamma(N)$ induces rapid growth of $T_{\cal{R S}}$ that saturates when $T_{\cal{S S}}$ reaches zero. Non-zero value of $T_{\cal{R S}}$ at the end of inflation plays crucial role in generation of non-Gaussianity parameter as we will see below.

In order to use \eqref{fNLviaIhorizon} one has to find $(\nabla T_{\cal RS})_{\parallel *}$, $(\nabla T_{\cal RS})_{\perp *}$ and $\tilde{M}_{\perp\perp *}$, $\tilde{M}_{\perp\parallel *}$. It is worth noting that the gradient of the transfer function at horizon crossing was computed analytically in the case of flat field-space metric in SRST approximation in \cite{Peterson:2010mv}. However, for arbitrary turn-rate the computation gets much more involved. Therefore we compute the terms above semi-analytically. To evaluate $(\nabla T_{\cal RS})_{ *}$ we use the finite-difference method. First, we perturb initial conditions for the background trajectory in $\phi$ and $\chi$ directions at the moment of horizon crossing and solve background equations with perturbed initial conditions. After computing derivatives of $T_{\cal RS}$ at horizon crossing with respect to the field basis, we rotate it to the kinematical basis using $e^a_{\perp}, e^a_{\parallel}$ to find $(\nabla T_{\cal RS})_{\parallel *}$, $(\nabla T_{\cal RS})_{\perp *}$. To find $\tilde{M}_{\perp\parallel *}$, $\tilde{M}_{\perp\perp *}$ we use equation \eqref{Mtilde} and first compute $(\nabla_a\eta_b)_*$ using the same approach with finite difference method described above. The rest of the terms of $\tilde{M}_{ab}$ we evaluate at the moment of horizon crossing using the background trajectory. Finally, we use basis vectors of the kinematical basis to compute the perpendicular and parallel components of $\tilde{M}_{ab}$ at horizon crossing. Using this method we numerically confirm equations \eqref{MtildeViaEta}. Quantities such as $\eta_{\perp *}, \eta_{\parallel *}$ and $\epsilon_*$ are computed at the background level at horizon crossing using an unperturbed trajectory.

In order to find values of $I_i$ coefficients in \eqref{Ihorizon} at horizon crossing we first compute $ \alpha_*$, $\beta_*$ defined in \eqref{alphabetahorizon} using the PyTransport code. For parameters \eqref{params} we find $\alpha_*=0.11, \beta_*=0.996$. These values explicitly illustrate the deviation from SRST approximation with $P_{\cal R*}=P_{\cal S*}$ and $C_{\cal RS*}=0$ where $\alpha_*=1$, $\beta_*=0$. Together with the value of the transfer function at the end of inflation $T_{\cal RS}=2.68$ we find the following values for coefficients
\begin{alignat}{3}
  I_{1*} &= 0.09, \quad I_{3*}&=0.53,   \quad I_{5*}=0.22,&\\
  I_{2*} &=0.03,  \quad I_{4*}&=0.67,   \quad I_{6*}=0.17.&
\end{alignat}
At the same time, we find
\begin{alignat}{3}
(\nabla T_{\cal RS})_{\parallel *} &=& 0.046, \quad \tilde{M}_{\perp\parallel*}&=2.31,  \\
  (\nabla T_{\cal RS})_{\perp *} &=&16.09,  \quad \tilde{M}_{\perp\perp *}&=-0.006.
\end{alignat}

Finally, we insert all the above numbers together with $\eta_{\parallel *}=-0.0005$, $\eta_{\perp *}=0.276$ and $\epsilon_*=0.0069$ (that gives the resulting turn-rate $\eta_{\perp *}/\sqrt{2\epsilon_*}\simeq 2.35$)
into \eqref{fNLviaIhorizon} and compute $f^{\rm loc}_{\rm NL}$ to be
\begin{equation}\label{fNLnumber}
 f^{(4)}_{\rm NL}= -\frac{5}{6} \left(0.006 \, I_{1*} + 1.89\, I_{2*} + 0.004 \,I_{3*} - 2.35 \,I_{4*} - 0.015 \,I_{5*} + 2.3 \,I_{6*}\right)\simeq 0.93.
\end{equation}
One can see that the largest contributions to the non-Gaussianity parameter are coming from the terms involving $I_{4*}, I_{6*}, I_{2*}$ (in descending order) that have in front of them $\eta_{\perp *}/\sqrt{2\epsilon_*}$, $\tilde{M}_{\perp\parallel *}$ and $\sqrt{2\epsilon_*}(\nabla T_{\cal RS})_{\perp *}$ respectively. Particularly interesting that the fourth and the sixth term appear with an opposite sign.

\begin{figure}
    \centering
    \includegraphics[scale=0.47]{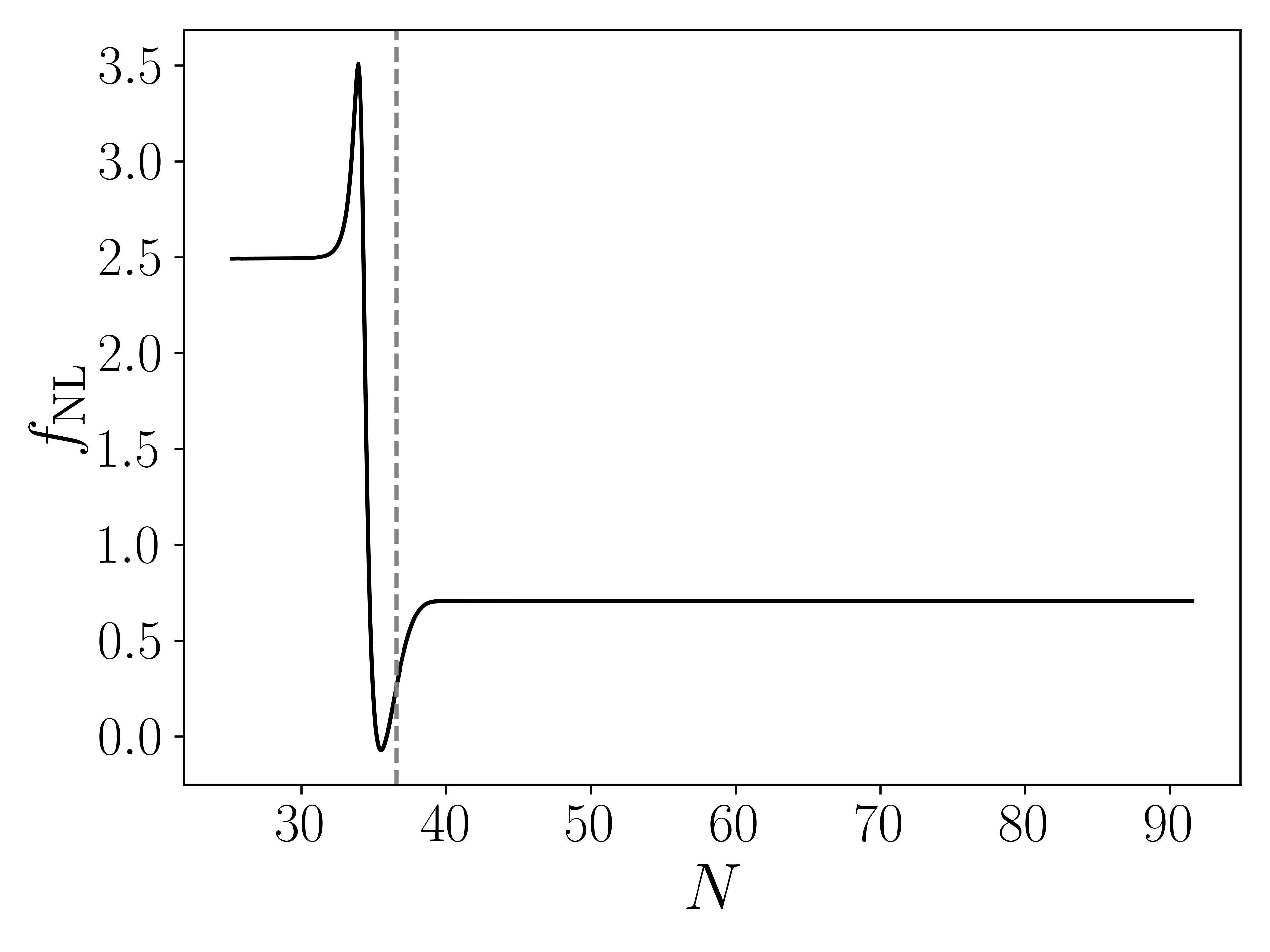}
    \caption{ Evolution of the non-Gaussianity parameter $f_{\rm NL}$ evaluated on the
equilateral triangle for the pivot scale $k$ that exits horizon 55 e-folds before the end of inflation.
    The parameter values used in the plot are given in Eq. \eqref{params}. The vertical gray dashed line represents the moment of horizon crossing $N_* = 36.6$. }
    \label{fig:fNL}
\end{figure}

\begin{figure}
    \centering
       \includegraphics[scale=1]{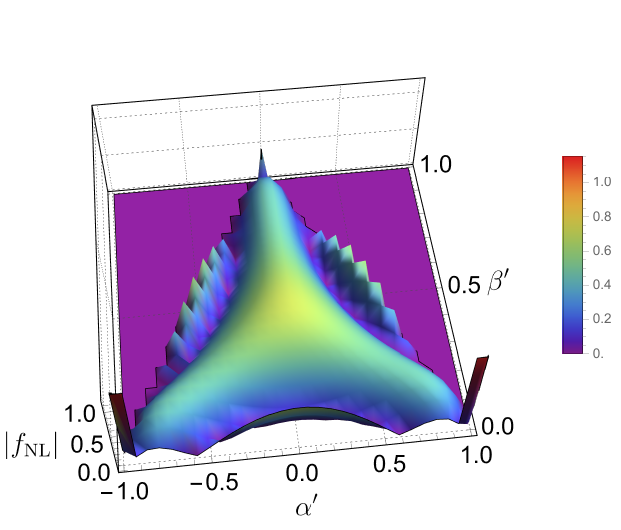}
    \caption{Shape dependence of $|f_{\rm NL}(\alpha', \beta')|$. }
    \label{plot:3D}
\end{figure}

\begin{figure}
    \centering
       \includegraphics[scale=0.7]{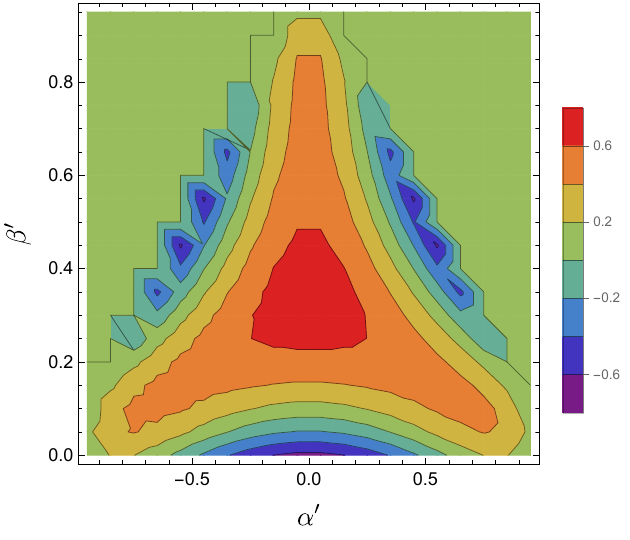}
       \includegraphics[scale=0.75]{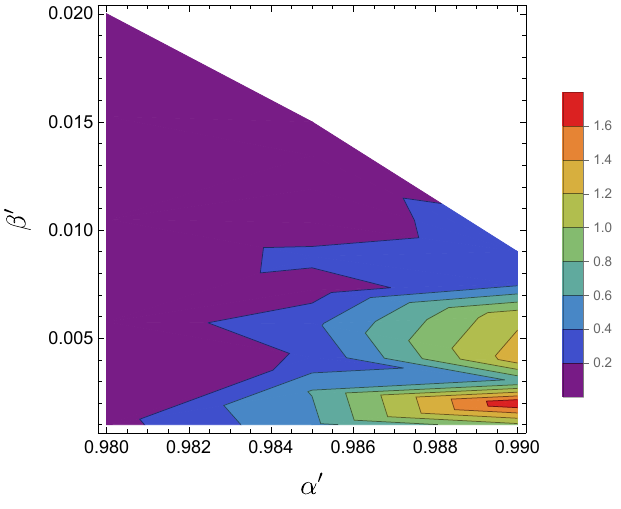}
    \caption{Left: The amplitude of $f_{\rm NL}(\alpha', \beta')$ zoomed in into a region $\alpha' \in (-0.8, 0.8)$ and $\beta\in (0,0.9)$. Right: The amplitude of $|f_{\rm NL}(\alpha', \beta')|$ zoomed into a squeezed triangles around 
    $(\alpha',\beta')= (0.99, 0)$ (right corner of Figure~\ref{plot:3D}).}
     \label{fig:contour0p8}
\end{figure}

In order to check our analytical result \eqref{fNLnumber} we compute $f_{\rm NL}$ numerically using the PyTransport code. The evolution of the non-Gaussianity parameter is shown in Fig. \ref{fig:fNL}. 
To find the bispectrum shape for all types of triangle configurations, we compute the amplitude of the bispectrum $f_{\rm NL}(\alpha', \beta')$ defined in Eq.~\eqref{fNLdef}
for $(\alpha', \beta')$ plane with vertices $(-1,0), (1,0)$ and $(0,1)$ respectively that parameterize any triangle shape based on the wave numbers
\begin{equation}
\begin{aligned}
 k_1&=\frac{k_*}{2}(1-\beta'),\\
    k_2&=\frac{k_*}{4}(1+\alpha'+\beta'),\\
    k_3&=\frac{k_*}{4}(1-\alpha'+\beta'),
\end{aligned}
\end{equation}
where $k_*$ is a pivot scale that exits horizon 55 e-folds before the end of inflation. The resulting shape dependence of $|f_{\rm NL}(\alpha', \beta')|$ is shown in Figure~\ref{plot:3D}. The bispectrum amplitude peaks in absolute value for equilateral configurations $k_1\sim k_2\sim k_3$
with $(\alpha', \beta')=(0,1/3)$, folded configurations $k_1+k_2\sim k_3$ with $(\alpha', \beta')=(0,0), (-1/2,1/2), (1/2,1/2)$ and local (or squeezed) configurations $k_1\ll k_2\sim k_3$ with $(\alpha', \beta')=(-1,0), (0,1), (1,0)$. In addition to that, the magnitude of the bispectrum amplitude has an opposite sign in equilateral and flattened configurations as shown in Figure~\ref{fig:contour0p8}, in agreement with ref.~\cite{Garcia-Saenz:2018ifx}\footnote{With an opposite sign of the bispectrum amplitude, due to an opposite sign in the definition \eqref{fNLdef}.}. In particular, for the current example we have $f^{\rm eq}_{\rm NL}\simeq - f^{\rm flat}_{\rm NL}=0.705$, where $f^{\rm eq}_{\rm NL}$ is a constant representing the amplitude of the equilateral shape and defined as \cite{Creminelli:2005hu}
\begin{equation}
    S^{\rm eq}=\frac{9}{10}f^{\rm eq}_{\rm NL}\left[-\left( \frac{k_1^2}{k_2k_3}+2\,perms.\right)+\left( \frac{k_1}{k_2}+5\,perms.\right) -2\right]=f^{\rm eq}_{\rm NL}\,  \tilde{S}^{\rm eq}.
\end{equation}
Similarly, a constant that represents the amplitude of the flattened shape, $f^{\rm flat}_{\rm NL}$, is \cite{Meerburg:2009ys}
\begin{equation}
    S^{\rm flat}=f^{\rm flat}_{\rm NL}\,  \tilde{S}^{\rm flat}=f^{\rm flat}_{\rm NL}\left( -\tilde{S}^{\rm eq} +\frac{9}{10}\right).
\end{equation}

From Figure~\ref{plot:3D}, we see that the amplitude of the bispectrum in the current example peaks in local, equilateral and flattened configurations, therefore a total shape may be written as a linear combination
\begin{equation}
    S=f^{\rm loc}_{\rm NL} \tilde{S}^{\rm loc}+f^{\rm eq}_{\rm NL} \tilde{S}^{\rm eq}+f^{\rm flat}_{\rm NL} \tilde{S}^{\rm flat},\label{Stotal}
\end{equation}
where a constant that defines the amplitude of local non-Gaussianity, $f^{\rm loc}_{\rm NL}$, is defined as
\begin{equation}
    S^{\rm loc}=\frac{3}{10}f^{\rm loc}_{\rm NL}\left( \frac{k_1^2}{k_2k_3}+2\,perms.\right)=f^{\rm loc}_{\rm NL} \tilde{S}^{\rm loc}.
\end{equation}
To compute the local non-Gaussianity parameter we use the definition of the bispectrum \cite{Renaux-Petel:2011rmu}
\begin{equation}
    B(k_1,k_2,k_3)=\frac{(2\pi)^4 {\cal P}^2 (k_*)S}{(k_1k_2k_3)^2}, \label{BviaS}
\end{equation}
where $S$ is the total shape function of the bispectrum and ${\cal P}(k_*)=2.1 \times 10^{-9}$. When the shape function of the bispectrum is local, the local non-Gaussianity parameter is 
\begin{equation}
   f^{\rm loc}_{\rm NL}=\frac{B(k_1,k_2,k_3)(k_1k_2k_3)^2}{(2\pi)^4 {\cal P}^2 (k_*)\, \tilde{S}^{\rm loc}},
\end{equation}
with wavenumbers $k_1, k_2, k_3$ fixed at local configurations. We have checked that at squeezed triangles $f^{\rm loc}_{\rm NL}$ coincides with $f_{\rm NL}(\alpha', \beta')$, the latter is shown on the right panel of Figure~\ref{fig:contour0p8}, and gives $|f^{\rm loc}_{\rm NL}|\sim {\cal O}(1)$ in agreement with the analytical result \eqref{fNLnumber}. 

Since in our case the total bispectrum shape is combined, it is worth checking the contribution from other shapes. From \eqref{Stotal} and \eqref{BviaS} one can find the local non-Gaussianity parameter in the form
\begin{equation}
   f^{\rm loc}_{\rm NL}\simeq\frac{1}{\tilde{S}^{\rm loc}}\left[\frac{B(k_1,k_2,k_3)(k_1k_2k_3)^2}{(2\pi)^4 {\cal P}^2 (k_*)}- f^{\rm eq}_{\rm NL} \left( 2\tilde{S}^{\rm eq}-\frac{9}{10}\right)\right], \label{fNLlocCombined}
\end{equation}
where we have used the general relation $\tilde{S}^{\rm flat}=\left( -\tilde{S}^{\rm eq} +\frac{9}{10}\right)$ and $f^{\rm eq}_{\rm NL}\simeq - f^{\rm flat}_{\rm NL}$ for the current example model. We have checked that the second contribution to \eqref{fNLlocCombined} does not spoil the prediction of $|f^{\rm loc}_{\rm NL}|\sim {\cal O}(1)$.

It is worth noting that the analytical result \eqref{fNLnumber} is derived in the approximation of single horizon crossing time of wavenumbers $k_1, k_2, k_3$, as we discuss in Section~\ref{subsec:SectionHorizon}. This corresponds to a near-equilateral momentum regime with $k_1\approx k_2\approx k_3$. In order to account for a highly squeezed momentum configuration $k_1\ll  k_2 \approx k_3$ one needs to take into account very different horizon exit times $t_1\ll  t_2 \approx t_3$. Extensive investigation of the highly squeezed limit significantly complicates the analytic treatment \cite{Kenton:2015lxa}. This goes beyond the scope of the current work and we leave it for the forthcoming explorations. Despite the limitations of single horizon crossing time approximation, our analytical result \eqref{fNLnumber} agrees well with numerical results for local non-Gaussianity parameter.

\section{Summary and discussion}\label{sec:Conclusion}

In this work, we have used the $\delta N$ formalism to derive a general analytical formula for the bispectrum generated on super-horizon scales in two-field, rapid-turn models of inflation. We have contrasted our result with the bispectrum computed in the multi-field SRST approximation \cite{Peterson:2010mv} and have demonstrated that a rapidly turning field trajectory and non-zero cross-correlation of field perturbations at horizon crossing generate novel additional contributions to the bispectrum. In general, the resulting bispectrum is not of the local shape. In order to explicitly identify potentially large model-independent contributions, we have expressed $f_{\rm NL}$ in terms of quantities evaluated either at horizon crossing or at the end of inflation. Our findings do not depend on the particular dynamics of models and are applicable to any two-field attractor model of inflation.

Additionally, we have illustrated our results through the explicit example of `angular inflation' \cite{Christodoulidis:2018qdw} and have demonstrated that contributions involving the curvature-isocurvature cross-correlation and the turn rate are dominant. We have found that the resulting bispectrum shape in the angular inflation model is a linear combination of local, equilateral and flattened shapes, with the same order of magnitude for their individual amplitudes. Despite the combined shape, we have verified numerically and analytically that, within a certain parameter range, this model can generate the local component $f^{\rm loc}_{\rm NL} \sim {\cal O}(1)$. 
In addition to the main discussion provided in this work, which focuses on the case when the modes exit the horizon during angular phase of inflation, we have also considered the case when the horizon exit happens during the (non-rapid-turn) radial phase within the same angular inflation model. In the radial phase the contribution from the turn-rate and the cross-correlation is negligible, and we have found that no significant amount of non-Gaussianity is generated. 
To provide further checks of our formula, we have also applied our result to realizations of orbital inflation \cite{Achucarro:2019pux} with negligible turn-rate and cross-correlation and have verified both analytically and numerically that the produced non-Gaussianity is very small, in agreement with the general prediction for $f_{\rm NL}$ in SRST approximation \cite{Peterson:2010mv}.

Several follow-up questions deserve further investigation. 
In this paper we have analyzed the general formula for the bispectrum in rapid-turn models of inflation for the case of single horizon crossing time of wave numbers. It was shown in \cite{Kenton:2015lxa} that accounting for different levels of hierarchy
between the scales can lead to substantial corrections to the bispectrum, of the order of $20\% $, in models of slow-turn inflation. It would be interesting to investigate (very) different horizon exit times in rapid-turn models of inflation. Another challenging question is the impact of different shape functions on the bispectrum, in particular those that give rise to
large non-Gaussianity. This could lead to new phenomenology and
provide potentially distinguishing observational probes. Last but not least, the generalization to the case with more than two fields would be interesting. We leave these exciting explorations for forthcoming work.

\section*{Acknowledgments}
We thank Ana Achúcarro, Ricardo Z. Ferreira,  Gonzalo A. Palma, Sébastien Renaux-Petel and Dong-Gang Wang  for stimulating discussions and comments on this work. The work of D.M. and G.S. is supported by the European Research Council under Grant No. 742104 and by the Swedish Research Council (VR) under grants 2018-03641 and 2019-02337. Nordita is supported in part by NordForsk. O.I. is grateful to the University of Leiden for hospitality.

\appendix

\section{Covariant definition for perturbations}\label{app:Covariant}
In this Appendix we describe a covariant formalism for studying perturbations in multi-field models. Each scalar field $\phi^a(x^{\mu})$ can be expanded around its classical background value $\varphi^a(t)$ to first order as
\begin{equation}
    \phi^a(x^{\mu})=\varphi^a(t)+\delta \phi^a(x^{\mu}).
\end{equation}
The fluctuation $\delta \phi^a(x^{\mu})$ defines a finite coordinate displacement from a classical trajectory, it is gauge-dependent and does not transform covariantly. To represent the field fluctuations in a covariant manner, a unique vector ${\cal Q}^a$ can be constructed. Up to third order in
fluctuations it relates to the field perturbation in the following way \cite{Gong:2011uw, Elliston:2012ab}\footnote{We use the version of the relation used in \cite{Garcia-Saenz:2019njm}.}
\begin{equation}
    \delta \phi^a= {\cal Q}^a-\frac{1}{2}\Gamma^a_{bc}{\cal Q}^b{\cal Q}^c+\frac{1}{6}\left(2\Gamma^a_{mn}\Gamma^n_{bc} -\Gamma^a_{bc,d}\right){\cal Q}^b{\cal Q}^c{\cal Q}^m+{\cal O}\left({\cal Q}^4 \right).
\end{equation}
Here $\Gamma^a_{bc}$ are the Christoffel symbols with respect to the
field space that are computed at background order in the fields. At the linear order  $\delta \phi^a$ and ${\cal Q}^a$ can be treated interchangeably. However, at higher orders, in particular for calculations of the three-point correlation function of
filed fluctuations,  the vector ${\cal Q}^a$ has to be used.
To ensure the gauge-invariance, it is convenient to use the gauge-invariant Mukhanov-Sasaki variables for the perturbations \cite{Mukhanov:1990me,Bassett:2005xm,Malik:2008im}
\begin{equation}
    Q^a={\cal Q}^a+\frac{\dot{\varphi}}{H}\psi.
\end{equation}
It is worth noting that in the spatially flat gauge vectors $Q^a$ and ${\cal Q}^a$ are interchangeable. Throughout the paper we use the gauge-invariant covariant definition of field perturbations, however in order to simplify a notation we denote it as $\delta \phi^a$.

\section{Full expressions for the bispectrum}
\label{app:FullexpressionHorizon}
In this Appendix we provide full expressions for the bispectrum found from \eqref{fNLGeneral}. The full result for the numerator of \eqref{fNLGeneral} expressed via 
the power spectrum at horizon crossing is 
\begin{equation}\label{FullNumeratorHorCrossing}
\begin{aligned}
&(2\epsilon_*)^{-1/2}\, N_a N_b N_{cd} K^{abcd}(k_1, k_2, k_3)=
 \\
 &{\cal P}_{{\cal R}_*}(k_1) {\cal P}_{{\cal R}_*}(k_2)
 \left[-\cos^2 \Delta_N T_{{\cal RS}} (\nabla T_{\cal RS})_{\parallel *}-\theta_{\parallel *} T_{{\cal RS}} -\frac{(\nabla \epsilon)_{\parallel *}}{2\epsilon_*}+\sin \Delta_N\cos \Delta_N (\nabla T_{\cal RS})_{\parallel *} \right]+\\
 &{\cal P}_{{\cal S}_*}(k_1) {\cal P}_{{\cal S}_*}(k_2)\left[
 \cos^2 \Delta_N T_{{\cal RS}}^2 (\nabla T_{\cal RS})_{\perp *}+ \theta_{\perp *}T_{{\cal RS}}^2-T_{{\cal RS}}^3\frac{(\nabla \epsilon)_{\perp *}}{2\epsilon_*}+\right.\\
 &\sin \Delta_N\cos \Delta_N T_{{\cal RS}}^3(\nabla T_{\cal RS})_{\perp *}
\biggr]+\\
  &{\cal C}_{{\cal RS}_*}(k_1) {\cal C}_{{\cal RS}_*}(k_2)\left[
  (1-T_{{\cal RS}}^2)(\theta_{\perp *}+\theta_{\parallel *} T_{{\cal RS}})-2T_{{\cal RS}}\frac{(\nabla \epsilon)_{\perp *}+(\nabla \epsilon)_{\parallel *} T_{{\cal RS}} }{2\epsilon_*}+\right.\\
  &\left((\nabla  T_{{\cal RS}})_{\perp *}+ \biggl.(\nabla  T_{{\cal RS}})_{\parallel *} T_{{\cal RS}}\right)\left(\cos^2 \Delta_N(1-T_{{\cal RS}}^2) + 2T_{{\cal RS}}\sin \Delta_N\cos \Delta_N \right) \biggr]+\\
   &{\cal P}_{{\cal R}_*}(k_1) {\cal P}_{{\cal S}_*}(k_2)\left[
   \cos^2\Delta_N T_{{\cal RS}}(\nabla  T_{{\cal RS}})_{\parallel *}+\theta_{\parallel *}T_{{\cal RS}}-T_{{\cal RS}}^2\frac{(\nabla \epsilon)_{\parallel *}}{2\epsilon_*}+
   \sin \Delta_N\cos \Delta_N T_{{\cal RS}}^2(\nabla  T_{{\cal RS}})_{\parallel *}
   \right]+\\
   &{\cal P}_{{\cal S}_*}(k_1) {\cal P}_{{\cal R}_*}(k_2)\left[
   -\cos^2\Delta_N T_{{\cal RS}}^2(\nabla  T_{{\cal RS}})_{\perp *}-\theta_{\perp *}T_{{\cal RS}}^2-T_{{\cal RS}}\frac{(\nabla \epsilon)_{\perp *}}{2\epsilon_*}+\right.\\
   &\sin \Delta_N\cos \Delta_N T_{{\cal RS}}(\nabla  T_{{\cal RS}})_{\perp *}
   \biggr]+\\
   &{\cal C}_{{\cal RS}_*}(k_1) {\cal P}_{{\cal R}_*}(k_2)\left[
   -T_{{\cal RS}}(\theta_{\perp *}+\theta_{\parallel *} T_{{\cal RS}})-\frac{1}{2\epsilon_*}\left((\nabla \epsilon)_{\perp *}+(\nabla \epsilon *)_{\parallel} T_{{\cal RS}}\right)+ \right.\\
   & \left((\nabla  T_{{\cal RS}})_{\perp *}+ 
   (\nabla  T_{{\cal RS}})_{\parallel *} T_{{\cal RS}}\right)\biggl.
   (-T_{{\cal RS}}\cos^2\Delta_N+\sin \Delta_N\cos \Delta_N)
   \biggr]+\\
   &{\cal P}_{{\cal R}_*}(k_1) {\cal C}_{{\cal RS}_*}(k_2)\left[
   (1-T_{{\cal RS}}^2)(\cos^2\Delta_N (\nabla  T_{{\cal RS}})_{\parallel *}+\theta_{\parallel *})-2T_{{\cal RS}}\frac{(\nabla \epsilon)_{\parallel *}}{2\epsilon_*}+\right.\\
   &2\sin \Delta_N\cos \Delta_N T_{{\cal RS}}(\nabla  T_{{\cal RS}})_{\parallel *}
   \biggr]+\\
   &{\cal C}_{{\cal RS}_*}(k_1) {\cal P}_{{\cal S}_*}(k_2)\left[
   T_{{\cal RS}}(\theta_{\perp *}+\theta_{\parallel *} T_{{\cal RS}})-T_{{\cal RS}}^2\frac{(\nabla \epsilon)_{\perp *}+(\nabla \epsilon)_{\parallel *} T_{{\cal RS}}}{2\epsilon_*}+\right.\\
  & \left((\nabla  T_{{\cal RS}})_{\perp *}+ 
   (\nabla  T_{{\cal RS}})_{\parallel *} T_{{\cal RS}}\right)\biggl.\left(T_{{\cal RS}}\cos^2\Delta_N+T_{{\cal RS}}^2\sin \Delta_N\cos \Delta_N \right)
   \biggr]+\\
   &{\cal P}_{{\cal S}_*}(k_1) {\cal C}_{{\cal RS}_*}(k_2)\left[
   T_{{\cal RS}}(1-T_{{\cal RS}}^2)(\cos^2\Delta_N(\nabla  T_{{\cal RS}})_{\perp *}+\theta_{\perp *})-2T_{{\cal RS}}^2\frac{(\nabla \epsilon)_{\perp *}}{2\epsilon_*}+\right.\\
   & 2\sin \Delta_N\cos \Delta_N T_{{\cal RS}}^2 (\nabla  T_{{\cal RS}})_{\perp *}
   \biggr]+ (\vec{k} {\rm \text { cyclic perms}}).
\end{aligned}
\end{equation}

The numerator of \eqref{fNLGeneral} expressed via
the power spectrum
at the end of inflation is given by
\begin{equation}\label{FullNumeratorEnd}
\begin{aligned}
&(2\epsilon_*)^{-1/2}   \, N_a N_b N_{cd} K^{abcd}(k_1, k_2, k_3)= \\
 &{\cal P}_{{\cal R}}(k_1) {\cal P}_{{\cal R}}(k_2)
 \left[-\frac{(\nabla \epsilon)_{\parallel *}}{2\epsilon_*}-\theta_{\parallel *}T_{{\cal RS}}
 \right]+\\
& \frac{{\cal C}_{{\cal RS}}(k_1) {\cal C}_{{\cal RS}}(k_2)}{T_{{\cal SS}}^2}\biggl[(1+T_{{\cal RS}}^2)(\theta_{\perp *}-T_{{\cal RS}}\theta_{\parallel *} )+((\nabla T_{\cal RS})_{\perp *}-T_{{\cal RS}}(\nabla T_{\cal RS})_{\parallel *} )
 \biggr]+\\
& \frac{{\cal P}_{{\cal R}}(k_1) {\cal C}_{{\cal RS}}(k_2)}{T_{{\cal SS}}}\biggl[\theta_{\parallel *}(1+T_{{\cal RS}}^2)+(\nabla T_{\cal RS})_{\parallel *}
 \biggr]+\\
 &\frac{{\cal C}_{{\cal RS}}(k_1) {\cal P}_{{\cal R}}(k_2)}{T_{{\cal SS}}}\left[-\frac{(\nabla \epsilon)_{\perp *}-(\nabla \epsilon)_{\parallel *} T_{{\cal RS}} }{2\epsilon_*}-T_{{\cal RS}}(\theta_{\perp *}-\theta_{\parallel *} T_{{\cal RS}})
 \right]+ (\vec{k} {\rm \text { cyclic perms}}).
 \end{aligned}
\end{equation}

\bibliographystyle{JHEP}
\bibliography{refs.bib}

\end{document}